\documentclass[11pt]{article}
\usepackage{tikz}
\usepackage{authblk}
\usepackage{geometry}
\usepackage[english]{babel}
\usepackage[utf8]{inputenc}
\usepackage[T1]{fontenc}
\usepackage{indentfirst}
\usepackage{amsmath}
\usepackage{amssymb}
\usepackage{amsthm}
\usepackage{proof}
\usepackage{eufrak}
\usepackage{graphicx}
\usepackage{psfrag}

\allowhyphens

\newcommand{\Pe}{\mathcal{P}}

\newcommand{\complex}{\mathbb{C}}

\newcommand{\valos}{\mathbb{R}}
\newcommand{\eps}{\varepsilon}

\newcommand{\ket}[1]{{\left|#1\right\rangle}}

\newcommand{\OO}{\mathrm{O}} 
\newcommand{\EE}{\mathrm{E}} 

\usepackage{ifpdf}

\ifpdf
\usepackage{epstopdf}
\usepackage[pdftex,colorlinks,urlcolor=blue,citecolor=blue,linkcolor=blue]{hyperref}
\else
\usepackage[hypertex,colorlinks,urlcolor=blue,citecolor=blue,linkcolor=blue]{hyperref}
\fi

\usepackage{tcolorbox}
\definecolor{aigray}{RGB}{180,180,180}
\definecolor{aibg}{RGB}{245,245,245}
\tcbuselibrary{skins,breakable}

\newtcolorbox{aiwritten}{
    enhanced,
    breakable,
    colback=aibg,
    frame hidden,
    borderline west={4pt}{0pt}{aigray},
    left=8mm,
    right=4mm,
    top=2mm,
    bottom=2mm,
    boxrule=0pt,
}

\usepackage{bm}

\newcommand{\e}{\mathrm{e}}

\newcommand{\vac}{\emptyset}

\usepackage{booktabs,array,tabularx,makecell,float,caption}
\usepackage{lmodern}
\usepackage{microtype}
\newcommand{\tr}{\mathrm{tr}}

\newcommand{\C}{\mathbb{C}}
\newcommand{\id}{\mathrm{Id}}
\newcommand{\signs}{\{+,-\}}
\newcolumntype{Y}{>{\raggedright\arraybackslash}X}

\newcolumntype{R}{>{\raggedleft\arraybackslash}p{2.0cm}}
\newcommand{\ii}{\mathrm{i}}
\newcommand{\ee}{\mathrm{e}}

\begin{document}
\numberwithin{equation}{section}

\title{
  Bethe Ansatz with a Large Language Model
}
\author[1]{Bal\'azs Pozsgay}
\author[1,2]{Istv\'an Vona}
\affil[1]{MTA-ELTE “Momentum” Integrable Quantum Dynamics Research Group,\protect\\
    ELTE E\"otv\"os Lor\'and University, Budapest, Hungary}
\affil[2]{Holographic Quantum Field Theory Research Group,\protect\\
    HUN-REN Wigner Research Centre for Physics, Budapest, Hungary}
\maketitle

\begin{abstract}
We explore the capability of a Large Language Model (LLM) to perform specific computations in 
  mathematical
physics: the task is to compute the coordinate Bethe Ansatz solution of selected integrable spin chain models. 
We select three integrable Hamiltonians for which the solutions were unpublished; two of the Hamiltonians are actually
new. We observed that the LLM semi-autonomously solved the task in all cases, with a few mistakes along
the way. These were corrected after the human researchers spotted them. The results of the LLM were checked against exact
diagonalization (performed by separate programs), and the derivations were also checked by the authors. The Bethe Ansatz
solutions are interesting in themselves. Our second model manifestly breaks left-right invariance, but it is
PT-symmetric, therefore its 
solution could be interesting for applications in Generalized Hydrodynamics. And our third model
is solved by a special form of the nested Bethe Ansatz, where the
model is interacting, but the nesting level has a free fermionic structure lacking $U(1)$-invariance. This structure
appears to be unique and it was found by the LLM.
We used ChatGPT 5.2 Pro and 5.4 Pro by OpenAI.
\end{abstract}

\section{Introduction and summary}

\label{sec:bev1}

Machine learning has been used for several years in multiple areas of science. More recently, Large Language Models
(LLMs) also started to appear as useful tools for making scientific discoveries. 
The progress is most apparent in mathematics.
Recent milestones include the new
solutions of selected Erdős’ problems and  other  problems selected by experts \cite{Erdos-problems,first-proof},
various discoveries using a combination of evolutionary 
programming and 
LLMs \cite{Tao-AlphaEvolve}, and displaying excellent performance in mathematical competitions such as the
International Mathematical Olympiad \cite{matharena}. 

In physics, machine learning has 
already been used in multiple research areas  \cite{machine-learning-physics},
and applications of the most recent versions of LLMs are starting to appear.
We do not attempt to review all contributions in this topic, instead we just mention a few selected papers 
\cite{LLM-qmb,cirac-llm,LLM-mandel,LLM-codes,LLM-earlyscience,LLM-claude-resummation,LLM-circuits}.

Given the success of LLMs in mathematics, it is a very natural idea to explore their usefulness in
mathematical physics.  The field of integrability is a specific area of mathematical physics, which is known for its proximity to
pure mathematics. 
Integrable models are special: they can be solved exactly. This means that many physical quantities
can be computed exactly, often in closed form expressions, without the need for approximations or numerical
simulations. The mathematical manipulations 
rely on special algebraic structures such as the Yang-Baxter relations \cite{jimbo-yb}.

Machine learning techniques have already been used in integrability
\cite{classical-Lax-pair-discovery-AI,R-matrixnet,AI-discovery-integrable-models-1} (and also for classical spin chains
\cite{LLM-Potts,LLM-Potts2}), however,  
to the best of our knowledge, the 
capabilities of LLMs have not yet been explored in a systematic way. There exist benchmarks for LLMs in mathematics
\cite{benchmark-frontiermath} 
and also in physics \cite{physbench1,physbench2}, but as far as we know, these benchmarks do not include
questions/problems in integrability.  This gives us motivation for our present study: we intend to explore the capabilities
of LLMs in research-level questions in integrability. 

In this paper we focus on a hallmark task of this field: computing the Bethe Ansatz solution of integrable spin chain models.
This involves constructing the exact coordinate space wave function of selected
interacting and integrable quantum spin 
chains; the method goes back to the solution of the Heisenberg spin chain by Hans Bethe in 1931 \cite{Bethe-XXX}.
Our goal is to explore to what extent an LLM can provide the Bethe Ansatz solution of a model, for which the solution
was not yet published in the literature.

We use ChatGPT 5.2 Pro and 5.4 Pro to perform the computations, and we
 also use the LLM to contribute to the text of this
manuscript.
{\bf Sections \ref{sec:bev1} and \ref{sec:bev2} are written exclusively by the human authors, but the other Sections
are mostly AI written, with human supervision. This includes overall editing and making various modifications.}

The structure of this paper is as follows. Below we summarize our key findings; we intend this summary for both experts
and  non-specialists. Afterwards we discuss  the potential checks of 
the output of the AI and the types of mistakes that the AI produced. In Section \ref{sec:bev2} we give a technical
description of the 
models to be solved, and in Section \ref{sec:CBA} we summarize the key ideas behind the Bethe Ansatz. Afterwards, in Sections
\ref{sec:Y1}-\ref{sec:freefermions} we solve the three selected models (the solution of the third model is presented in two
Sections). Finally, we present numerical data for the predicted spectra in the Appendices
\ref{app:Y1}-\ref{app:Y3}. 

An overall outlook is given earlier at the end of the 
Introduction, in Subsection \ref{sec:outlook}.

We also publish the output of the LLM for a selected example prompt; this can be found as an ancillary material added to
this document. This output concerns the solution of one of our models, but it is not the full solution and it does not
coincide with the corresponding parts from the main text.
The actual material of this paper was compiled using
several calls to the LLM, and the ancillary material serves as an illustration of the work process. 

\subsection{A non-technical summary of this work}

We used ChatGPT 5.2 Pro and 5.4 Pro from OpenAI. We instructed this LLM to find solutions to three integrable spin chain
models, 
whose Bethe Ansatz solution is not available in the literature. 

We chose three spin chain models that we named Y1, Y2 and Y3. Below we describe the three models, and give a few remarks
about  
their physical properties and the difficulty level associated with their solution. Assessing this difficulty level
involves personal judgement. However, we added these remarks so that non-specialists
could have a rough assessment of the results. The concrete definition of the three models (together with some additional
technical details) will be given later in Section
\ref{sec:bev2}. 

\begin{itemize}
\item {\bf Model Y1} is a relatively simple model, which is related to the XXZ Heisenberg chain. The connection is not
  immediately obvious, but once it is understood, the solution of the model becomes very simple. We instructed the LLM
  to solve the model {\it without} prompting it to find connections to known models.

Computing the Bethe Ansatz for this model is a good learning exercise for an undergraduate student. In contrast, the
connection to the XXZ chain can be overlooked, and even experts might miss it.
\item {\bf Model Y2} is a previously unpublished model. It is physically interesting, because it breaks
  left-right reflection invariance at the macroscopic level. It has two types of excitations over the pseudovacuum,
  therefore, we expected that it can be solved by the so-called {\it nested Bethe Ansatz}.

The solution of this model could be a project for an MSc student, or perhaps a learning exercise for a PhD student in
the first year. In this paper we do not compute the transport properties of the model, but we believe that
the model could be interesting for Generalized Hydrodynamics due to the explicit 
breaking of space reflection invariance.

\item {\bf Model Y3} was published earlier in \cite{sajat-medium}, but its solution was not given there. It is an
  $SU(2)$-symmetric spin chain model with 4-site interactions. Simple arguments reveal that the model has two types of
  excitations over the pseudovacuum, therefore, we  expected a solution via the nested Bethe Ansatz. However, this
  time there
  is no $U(1)$-symmetry at the nesting level, which makes the model unusual. In fact, the model turns out to have an
  8-vertex type $R$-matrix at the nesting level. Furthermore, this particular $R$-matrix is of the free fermion type.
This free fermionic property was found by the LLM and it was  a surprise
  to us.

The difficulty level of this solution corresponds to an advanced project for a PhD student, or perhaps a learning
exercise for an early postdoc.
\end{itemize}

Now we summarize the results and our experiences:

\begin{itemize}
\item The LLM was able to compute the Bethe Ansatz solution of all these models. Notably, it recognized the free
  fermionic structure of model Y3, which makes its solution possible despite the lack of $U(1)$-invariance on the nesting
  level. Furthermore, it discovered a simple eigenvector
  for the auxiliary transfer matrix on the nesting level, which eventually leads to simple but non-trivial Bethe
  equations and Bethe states, at least for a subset of the eigenstates of the model. Therefore, we believe that these
  results would deserve publishing if they were derived by a human: they appear as interesting additions to the theory
  of integrable models.
\item The LLM makes mistakes, but it is able to correct them, once prompted. Therefore, intermediate and final
  results cannot be trusted without independent checks. We discuss this in more detail in Subsection \ref{sec:checks} below.

\item The LLM writes and runs python scripts to check certain parts of the computation.
   An independent reading and running the scripts can be useful for assessing the correctness
  of the solution. Hallucinations can appear in the output of the LLM even after it wrote and ran correct scripts, see
  Subsection \ref{sec:checks} below. 
\item The LLM is able to summarize the solution and present it in a form which fits the style and requirements of a
  scientific journal.
\item There is a dramatic difference in the performances of the free and Pro versions of ChatGPT 5.2 and 5.4. In the case
  of model Y3 we ran the same prompts also for the free versions, and the LLM was not able to compute the Bethe Ansatz
  solution. In fact, it did not make any useful steps.
\end{itemize}

\subsection{Checking the output of the AI}

\label{sec:checks}

Our specific problems are such that they are theoretical in nature, but can be checked easily in concrete examples.

We aim at computing the spectrum of integrable Hamiltonians in a finite volume. The energy eigenvalues are computed from
the so-called Bethe rapidities, which are solutions to the Bethe equations. 
They form a set of coupled non-linear equations in several
variables. In small system sizes it is relatively easy to find concrete numerical solutions, from
which the corresponding energy eigenvalue can be computed. These numerical values can then be compared to exact
diagonalization.

Mistaken Bethe equations almost always lead to incorrect energy
eigenvalues. Therefore, a numerical check of the energies for a few randomly selected solutions is typically enough to
judge the correctness of the solution. This procedure has minor time cost: the numerical solution of the
Bethe equations is performed by the LLM (it can be checked separately by us), and exact diagonalization in small volumes
is typically enough to judge the correctness. 

Interestingly, we found that ``hallucination'' of the LLM can appear in the numerics too. We observed runs where we
instructed the LLM to compare its numerical predictions (from the Bethe equations) with exact diagonalization. In such a
case the LLM writes python scripts, it runs the scripts, and analyzes the outputs. We found that the LLM sometimes
hallucinated the numbers for the energy eigenvalues; those values were actually not there in the spectrum. 
In our experiments, the hallucinations appeared only in those cases when there was a mistake in the Bethe Ansatz
computation itself, and the 
correct numerics would have uncovered the mismatch. 
After such 
occurrences we 
decided to perform the exact diagonalization by a separate program in each case.

\subsection{Types of mistakes of the LLM}

We found multiple types of mistakes in the analytical calculations of the LLM. Examples include:
\begin{itemize}
\item Sometimes the LLM assumed that a certain model was solvable by a simple Bethe Ansatz, and it extended Bethe
  equations 
  from the two-particle case to the general $M$-particle problem, without checking all the conditions for the
  correctness of the Ansatz. Once confronted with the mismatches in the numerical data, it realized the mistake and it
  went on to compute the nested Bethe Ansatz.
  We never had to explicitly instruct the LLM to carry out nested Bethe Ansatz for any of the models.
\item  Sometimes minor mistakes were made. Interestingly, these mistakes were such that they could easily occur
  also with a human researcher, before finalizing all conventions and performing all possible checks. For example, a
  typical mistake was that in 
  the Bethe 
  equations the $S$-matrix had a wrong order of the momenta,
  or the LLM would use the
  so-called checked R-matrix $\check R(u,v)$ instead of the standard one, or there was an extra minus sign in the Bethe
  equations, etc. 
\item Sometimes intermediate formulas had mistakes, even though the final formulas were correct.
\end{itemize}

\subsection{Discussion and outlook}

\label{sec:outlook}

\paragraph{The solution of model Y3.}

Using the LLM we computed the full solution of model Y3, to be presented below. This model appears
interesting due to its unique Bethe Ansatz structure: it is interacting, solvable by nested Bethe Ansatz, where the
nesting level has free fermions without $U(1)$-symmetry. Our solution  produces
all eigenvectors of the second level Bethe Ansatz and likely most eigenvectors of the actual model. We did not
investigate the completeness of the Bethe Ansatz.

It might be that there is a simpler solution than ours. The current solution is well adapted for numerical studies, but
inconvenient for taking the thermodynamic limit. Therefore, alternative formulations are also desirable.

\bigskip

\paragraph{General outlook.}

The LLM displayed an excellent performance on the selected  research-level questions. The difficulty level
was such that they could be good projects for a PhD student, and the results could have been published if there was no
AI involved. There is considerable gain also for an expert, due to the hugely reduced time in working out the details of
the solution.

It would be interesting to attack other problems in the field of integrable models, with increasing difficulty levels. It
is possible that solutions will be found for open problems that are difficult even for the experts. 

All of this progress calls for automated verification of the results of the LLM.
In our concrete problems a quick check was
possible by comparing to exact diagonalization. However, such an easy route might not always be available. It is
desirable to have  proof-checking mechanisms, similar to those developed in mathematics, see for example
\cite{aristotle}. For recent contributions towards this goal in physics and computer science see
\cite{Lean4Physics,physprover,merlean}. 

Finally we mention the question of benchmarking AI using open problems. There are online repositories dedicated to
open problems, both in mathematics and in physics, and there are various initiatives to build benchmarks  (see for
example \cite{erdosproblems,unsolvedmath,solve,IMProofBench,first-proof,physbench1,physbench2}). It is our impression
that mathematical physics is underrepresented in these lists. In particular, problems from the field of integrable
models have not yet been added.
This research area can be seen as having overlaps with both physics and pure mathematics, therefore, 
it could be useful to add problems from this field to the benchmarks.

\section{Integrable spin chain models}

\label{sec:bev2}

We consider finite spin chains with local dimension 2, therefore the Hilbert space is $\otimes_{j=1}^L \complex^2$.

We treat translationally invariant local Hamiltonians:
\begin{equation}
  H=\sum_{j=1}^L h(j)
\end{equation}
Here $h(j)$ is the operator density localized around the site $j$. We always treat local models, which means that $h(j)$
is a short range operator which spans a finite number of sites. In the concrete examples the range
will be 3 and 4. We always treat periodic boundary conditions, because in the typical case this choice leads to the
simplest possible solutions of the models.

There are various definitions and diagnostic criteria of integrability \cite{caux-integrability}. We will work with the
definition that a local spin chain is integrable, if there exists a family of charges $Q_\alpha$, all of them with local
operator densities, which form a commuting family such that the Hamiltonian is a member of the family. We will use a
convention such that the index $\alpha$ coincides with the range of the operator density of $Q_\alpha$; in this case the
possible values of $\alpha$ form a subset of the natural numbers.

A prominent example is the XXZ spin chain, whose Hamiltonian density can be written as
\begin{equation}
  \label{XXZH}
  h(j)=X_jX_{j+1}+Y_jY_{j+1}+\Delta Z_jZ_{j+1}
\end{equation}
Here we used the notation $X, Y, Z$ for the Pauli operators, and $\Delta\in\valos$ is the anisotropy parameter.

In the case of $\Delta=1$ the model is the XXX chain, whose solution was computed by Hans Bethe \cite{Bethe-XXX}. Bethe
proposed an Ansatz for the explicit coordinate space wave function, whose validity can be confirmed by direct
computations. The anisotropic chain was first solved in \cite{XXZ1}.

Now we summarize simple commutativity relations that guarantee the integrability of spin chain models.

The integrability of the nearest neighbour interacting spin chain models can be established by the Algebraic Bethe
Ansatz methods, which ultimately rely on particular solutions to the Yang-Baxter equations
\cite{faddeev-how-aba-works}. However, integrability can be checked more easily via the so-called Reshetikhin
condition. The statement is that if a Hamiltonian is derived from a solution of the Yang-Baxter relation, then we have a
commutativity
\begin{equation}
  \label{Resh}
  [H,Q_3]=0
\end{equation}
where 
\begin{equation}
  H=\sum_j h(j),\qquad Q_3=\sum_j   q_3(j),
\end{equation}
where
\begin{equation}
  q_3(j)=i[h(j),h(j+1)]+\tilde h(j),
\end{equation}
and $\tilde h(j)$ is a two-site operator density. Both $h(j)$ and $\tilde h(j)$ originate from derivatives of the
$R$-matrix, which solves the YB relations. The precise connections will not be used here.

Recently a considerable amount of work was devoted to find theorems for the opposite direction: proving that an infinite
tower of charges exist, if there is a $h(j)$ and $\tilde h(j)$ that satisfy the above commutativity relation
\cite{dichotomy,resh-cond-proof}.  

An extension of this framework to Hamiltonians with multi-site interactions was presented in \cite{sajat-medium}. A
generalization of the Reshetikhin condition was given as follows. Consider a translationally invariant Hamiltonian with
a three-site interaction  $h(j)$. Then the condition is that
\begin{equation}
  [H,Q_5]=0,
\end{equation}
where $Q_5$ is an extensive 5-site operator with its density being
\begin{equation}
  q_5(j)=i[h(j),h(j+1)+h(j+2)]+\tilde h(j),
\end{equation}
where $\tilde h(j)$ is another three-site operator. The extension to longer interaction ranges follows along the same lines.

It was conjectured in \cite{sajat-medium} that most solutions of these conditions actually lead to a medium range
integrable model with a tower of conserved charges. The precise conditions for the validity of this conjecture have not
been established yet. However, this generalized condition was used successfully in  \cite{sajat-medium}  to find new
integrable models, and an example is our model Y3 treated below. Furthermore,  we found the unpublished model Y2 using
the same formalism. 

All of this means that the integrability of the models Y2 and Y3 is not yet rigorously established, because the
corresponding solutions of the Yang-Baxter relations have not yet been found. However, the existence of the Bethe
Ansatz, together with a numerical check of the concrete predictions for the energy eigenvalues up to 4-particle level
strongly suggests that the models are indeed integrable.

\subsection{Our models}

We consider three models, for which the Bethe Ansatz solution has not yet been presented in the literature.

In interpreting and solving the models we will use the following terminology: the state with all spins up is called the
reference state (or pseudovacuum). Down spins embedded into a sea of up spins can be seen as the excitations or
particles. 

\bigskip
{\bf Model Y1}
\bigskip

The first model  is given by
\begin{equation}
  \label{Y1H}
  H_{Y1}=\sum_{j=1}^L h_{Y1}(j)
\end{equation}
The model has a three-site density, which is most conveniently written as
\begin{equation}
  \label{Y1hh}
  h_{Y1}(j)=Y_{j}Z_{j+1}X_{j+2}-X_{j}Z_{j+1}Y_{j+2}+\Delta (Z_{j}+Z_{j+3})(X_{j+1}X_{j+2}+Y_{j+1}Y_{j+2})
\end{equation}
In this representation the operator density spans 4 sites, but each term is a three-site operator. $\Delta$ is a real
coupling constant.

We believe that this particular Hamiltonian has not yet appeared in the literature. However, there is a direct
connection to the XXZ chain. To see this connection, consider the Hamiltonian
 $H_{tw}=\sum_{j=1}^L h_{tw}(j)$, where
\begin{equation}
  \label{XXZtwH}
  h_{tw}(j)=X_jY_{j+1}-Y_jX_{j+1}+\Delta Z_jZ_{j+1}
\end{equation}
This model can be seen as a twisted version of the XXZ chain. More precisely, it is unitarily equivalent to the model
given by \eqref{XXZH} in every finite volume $L$ which is a multiple of 4. In other volumes it can be seen as the XXZ
chain with twisted boundary conditions.

It turns out that $H_{Y1}$ from \eqref{Y1hh} can be written as
\begin{equation}
H_{Y1}=-\frac{i}{2}\sum_{j=1}^L [h_{tw}(j),h_{tw}(j+1)]
\end{equation}
and direct computations in finite volumes show that
\begin{equation}
  [H_{Y1},H_{tw}]=0
\end{equation}
In other words, $H_{Y1}$ is simply the first higher conserved charge for the nearest neighbour model given by
$H_{tw}$. In particular, the Reshetikhin condition \eqref{Resh} is satisfied with $\tilde h(j)=0$.
The operator $h_{Y1}(j)$ can also be interpreted as the energy current in the model $H_{tw}$.

All of this implies that the eigenstates of $H_{Y1}$ will coincide with those of $H_{tw}$. For $H_{tw}$ the standard
Bethe Ansatz applies, with minor differences due to the twist in the hopping terms.

This explains why solving $H_{Y1}$ is a good first problem for the LLM: the Bethe Ansatz eigenstates have a
relatively simple structure, nevertheless the 
Hamiltonian itself appears new, and it is unlikely that an LLM would discover the connections, unless specifically prompted.

When asked to find the exact eigenstates, the LLM did not mention this connection. However, when we asked if there is
any connection between $H_{Y1}$ and any known model in the literature, it found the connection.

\bigskip
{\bf Model Y2}
\bigskip

This is a Hamiltonian which was earlier discovered by the authors but remained unpublished so far. It is given by
\begin{equation}
  \label{Y2H}
  H_{Y2}=\sum_{j=1}^L h_{Y2}(j)
\end{equation}
with
\begin{equation}
h_{Y2}(j)=  \sigma_{j}^{+}D_{j+1}(\gamma)\sigma_{j+2}^{-}+\sigma_{j}^{-}D_{j+1}(-\gamma)\sigma_{j+2}^{+}+\frac{\sin(2\gamma)}{2}(Z_{j}Z_{j+1}-1),
\end{equation}
where $\gamma\in [0,\pi)$ and
\begin{equation}
  D(\gamma)=e^{\ii\gamma Z}=
  \begin{pmatrix}
    e^{\ii\gamma} & \\ & e^{-\ii\gamma}
  \end{pmatrix}
\end{equation}
For generic $\gamma$ this model breaks the left-right reflection symmetry which is usually present in integrable
models. However, space reflection combined with complex conjugation is an invariance of the model, therefore we say
the model is PT-symmetric.

For $\gamma=0$  the model
can be seen as the union of two XX models, because particles can freely hop by two sites, leading to a complete dynamical
separation of the even 
and odd sub-lattices. This also means that there are two $U(1)$ symmetries in the model, those corresponding to the two
sublattices. 

Turning on $\gamma\ne
0$ the particles can still hop only within each sub-lattice, the two $U(1)$ symmetries survive, however, we obtain an
interaction term between the sublattices.

This dynamical picture suggests that the model is solvable by a nested Bethe Ansatz, where two excitation types
correspond to particles propagating on the individual sub-lattices. 

\bigskip
{\bf Model Y3}
\bigskip

This model is given by
\begin{equation}
  \label{Y3H}
  H_{Y3}=\sum_{j=1}^L h_{Y3}(j)
\end{equation}
with
\begin{equation}
  h_{Y3}(j)=(\Pe_{j,j+3}-1)(\Pe_{j+1,j+2}-1)-(\Pe_{j,j+2}-1),
\end{equation}
where $\Pe_{j,k}$ is the permutation operator acting on two qubits. The Hamiltonian only involves operators that are
symmetric with respect to a global $SU(2)$ symmetry, therefore the model itself is rotationally symmetric. We also note that
$\Pe_{j,k}-1$ is proportional to the projector to the $SU(2)$-singlet prepared on sites $j$ and $k$.

The model was published in the work \cite{sajat-medium}, where the authors classified all $SU(2)$-symmetric, space
reflection symmetric and translationally invariant Hamiltonians with 4-site interactions. In fact, this was the only new
model that appeared in the classification once simple variants of the XXX model were excluded. The Bethe Ansatz solution
of the model was not given in \cite{sajat-medium}, and none of us considered the problem since the publication of
\cite{sajat-medium}.

Simple arguments reveal that this model should be solvable by the nested Bethe Ansatz. The reference state is an
eigenstate with zero energy. The propagation of isolated particles is generated only by the term including
$\Pe_{j,j+2}$, because the first product acts identically as zero on all configurations where a single down spin is embedded
into a sea of up spins. It follows that isolated particles can hop by two sites only, therefore, they will propagate on
the odd and even sub-lattices. We obtain that once again there will be two types of excitations. The first term acts in
a non-trivial way only if two excitations come close to each other. However, there is a crucial difference here as
opposed to the model Y2: the interaction terms are such that they can move particles between the
sub-lattices. Therefore, this model does not conserve ``excitation type''.

Based on this simple analysis we expect a non-trivial nested Bethe Ansatz solution, where we cannot expect to find
``particle conservation'' on the nesting level.

\section{Coordinate Bethe Ansatz -- A summary}
\label{sec:CBA}

The coordinate Bethe Ansatz solves a one-dimensional many-body eigenvalue problem by writing the wave function
separately in each region with fixed particle order. In such an ordered sector the excitations propagate freely, so the
wave function is a superposition of plane waves; all non-trivial information is then concentrated in the contact
conditions, where two particles exchange their order. In an integrable model these local matching conditions are
sufficient to reconstruct the full many-body state, because the scattering factorizes into two-body processes. This
point of view goes back to Bethe's solution of the spin-$\tfrac12$ chain and was later extended to continuum gases and
multicomponent systems by Lieb, Liniger, Yang, and Sutherland; textbook accounts may be found in
Refs.~\cite{gaudin-book-forditas,Korepin-book}. Throughout this section we denote by $x_1<\cdots<x_N$ the ordered
particle positions, by $L$ the period, by $z_j$ a special choice of rapidity variables (often $z_j=\mathrm{e}^{\ii p_j}$
where $p_j$ is the lattice momentum), and by $\varepsilon(z)$ the one-particle dispersion. 

\subsection{Scalar coordinate Bethe Ansatz}

When each excitation carries no additional internal label, the standard Ansatz on the ordered sector is
\begin{equation}
\psi(x_1,\dots,x_N)=\sum_{P\in S_N} A(P)\prod_{a=1}^N z_{P_a}^{x_a},
\end{equation}
where $S_N$ is the permutation group on $N$ objects. Here $A(P)$ are scalar amplitudes, depending on the permutation. 

If all separations are larger than the interaction range, the Schr\"odinger equation reduces to a sum of one-particle problems and fixes the energy in additive form,
\begin{equation}
E=\sum_{j=1}^N \varepsilon(z_j).
\end{equation}
The remaining information comes from the contact equations. Solving the two-body problem yields a scalar scattering factor $S(z_a,z_b)$, and in the $N$-particle sector one obtains the exchange rule
\begin{equation}
A(\dots,b,a,\dots)=S(z_a,z_b)\,A(\dots,a,b,\dots).
\end{equation}
Because $S(z_a,z_b)$ is a number rather than a matrix, different sequences of pairwise exchanges automatically give the same amplitude. After fixing one reference coefficient, for instance $A(1,2,\dots,N)=1$, every other amplitude is therefore obtained as the product of the two-body factors associated with the inversions of the permutation $P$.

Periodic boundary conditions are imposed by taking one particle once around the ring. With the exchange convention above, the resulting Bethe equations are
\begin{equation}
z_j^L=\prod_{\substack{k=1\\k\ne j}}^N S(z_k,z_j),
\qquad j=1,\dots,N.
\end{equation}
If the rapidities are written as $z_j=\mathrm{e}^{\ii p_j}$, this is equivalently a quantization condition for the
quasi-momenta $p_j$. The special case $S=-1$ gives the familiar free-fermion quantization $z_j^L=(-1)^{N-1}$.

\subsection{Nested coordinate Bethe Ansatz}

The scalar construction is not sufficient when each excitation carries an internal label, such as a flavor, a sublattice index, or a sign variable $\sigma\in\{+,-\}$. The coordinate part of the wave function must then be accompanied by an internal amplitude. A convenient form is
\begin{equation}
\Psi(x_1,\dots,x_N)=\sum_{\pi\in S_N}\;\sum_{\alpha_1,\dots,\alpha_N}
A_{\pi}^{\alpha_1\dots\alpha_N}
\prod_{j=1}^N \phi_{\alpha_j}(z_{\pi(j)};x_j),
\end{equation}
where $\alpha_j$ labels the internal state. In the simplest cases one has $\phi_{\alpha}(z;x)=z^x$. In cases when the
internal label corresponds to two sub-lattices,  sometimes
the sub-lattice index can be conveniently encoded by a sign and one may take $\phi_{\sigma}(z;x)=(\sigma z)^x$.

The bulk equation again fixes an additive energy,
\begin{equation}
E=\sum_{j=1}^N \varepsilon(z_j),
\end{equation}
but the contact equations now relate vectors rather than scalars. If $A_{\pi}$ denotes the internal amplitude vector attached to the permutation $\pi$, then the exchange of neighboring rapidities is governed by a braided two-body matrix,
\begin{equation}
A_{\pi s_j}=\check R_{j,j+1}\bigl(z_{\pi(j)},z_{\pi(j+1)}\bigr)A_{\pi},
\qquad j=1,\dots,N-1.
\end{equation}
Here $s_j$ is the adjacent transposition exchanging $j$ and $j+1$. Factorized many-body scattering requires the
corresponding unbraided matrix $R=P\check R$ to satisfy the Yang-Baxter equation 
\begin{equation}
R_{12}(u,v)\,R_{13}(u,w)\,R_{23}(v,w)
=
R_{23}(v,w)\,R_{13}(u,w)\,R_{12}(u,v),
\end{equation}
If the Yang-Baxter relations are satisfied, the wave function is well defined, and it can be an exact eigenstate
(formally in infinite volume) if all contact terms in the Hamiltonian are compatible with the factorized form. This
needs to be checked in all cases.

Periodic boundary conditions now become an eigenvalue problem in the internal space. Introducing the monodromy and
transfer matrices using an auxiliary space,
\begin{equation}
T_a(u)=R_{aN}(u,z_N)\cdots R_{a1}(u,z_1),
\qquad
 t(u)=\tr_a T_a(u),
\end{equation}
the Yang-Baxter relation implies $[t(u),t(v)]=0$.

The $R$-matrix satisfies the so-called regularity condition, which in our cases takes the form
\begin{equation}
  \check R(u,u)=-1
\end{equation}
After specializing $u$ to one of the physical rapidities and using the
regularity condition
 one obtains the specialized operators
\begin{equation}
\mathcal T_k(\{z\})=-t(z_k)
=R_{k,k-1}(z_k,z_{k-1})\cdots R_{k1}(z_k,z_1)
 R_{kN}(z_k,z_N)\cdots R_{k,k+1}(z_k,z_{k+1}),
\end{equation}
with the convention that an empty product equals the identity.
These operators  implement the winding of particle $k$ around the ring.

The periodicity conditions can then be written as
\begin{equation}
z_k^L\,\mathcal T_k(\{z\})A=A,
\qquad k=1,\dots,N,
\end{equation}
with $A$ the common internal amplitude vector. If $A$ is a simultaneous eigenvector of these commuting operators, with eigenvalues $\Lambda_k(\{z\})$, the nested Bethe equations take the compact form
\begin{equation}
z_k^L\,\Lambda_k(\{z\})=1,
\qquad k=1,\dots,N.
\end{equation}

The remaining step is the solution of the auxiliary transfer-matrix problem. If the internal $R$-matrix can be
normalized to a six-vertex form and admits a pseudovacuum, one solves this auxiliary problem by a second Bethe Ansatz,
by using its algebraic form \cite{fedor-lectures}.
In other situations different techniques need to be applied; below our model Y3 will show such an unusual behaviour.

\subsection{Conventions for the spin-$1/2$ chains}

In the Sections below we will use the following conventions.

Let $\sigma_j^{\pm}=(X_j\pm \ii Y_j)/2$, let $|\vac\rangle=|\uparrow\cdots\uparrow\rangle$, and define ordered basis states
\begin{equation}
|x_1,\dots,x_N\rangle=\sigma^-_{x_1}\cdots\sigma^-_{x_N}|\vac\rangle,
\qquad 1\le x_1<\cdots<x_N\le L.
\end{equation}
The local spin operators act on chain sites modulo $L$, whereas the ordered coordinates $x_1<\cdots<x_N$ are treated on
the usual covering space; periodicity is imposed later through $\psi(x_1,\dots,x_N)=\psi(x_2,\dots,x_N,x_1+L)$.  

\section{Solution of model Y1}

\label{sec:Y1}

We rewrite the Hamiltonian \eqref{Y1H} as
\begin{equation}
\begin{aligned}
H=2\sum_{j=1}^L\Big[&\ii\big(\sigma_j^-Z_{j+1}\sigma_{j+2}^+-\sigma_j^+Z_{j+1}\sigma_{j+2}^-\big) \\
&+\Delta\big(Z_j(\sigma_{j+1}^+\sigma_{j+2}^-+\sigma_{j+1}^-\sigma_{j+2}^+)+(\sigma_j^+\sigma_{j+1}^-+\sigma_j^-\sigma_{j+1}^+)Z_{j+2}\big)\Big],
\end{aligned}
\end{equation}
We see explicitly that the number of particles (down spins) is conserved.

In the one-magnon sector one finds directly
\begin{equation}
H|x\rangle=4\Delta\big(|x-1\rangle+|x+1\rangle\big)+2\ii|x-2\rangle-2\ii|x+2\rangle,
\end{equation}
which gives the plane-wave eigenvalue
\begin{equation}
\eps(z)=4\Delta(z+z^{-1})+2\ii(z^2-z^{-2}),\qquad \eps(p)=8\Delta\cos p-4\sin 2p\quad (z=\e^{\ii p}).
\end{equation}
In the ordered two-magnon sector, the bulk equation for $x_2\ge x_1+3$ is
\begin{equation}
\begin{aligned}
E\,\psi(x_1,x_2)={}&4\Delta\big[\psi(x_1-1,x_2)+\psi(x_1+1,x_2)+\psi(x_1,x_2-1)+\psi(x_1,x_2+1)\big]\\
&+2\ii\big[\psi(x_1+2,x_2)+\psi(x_1,x_2+2)\big]-2\ii\big[\psi(x_1-2,x_2)+\psi(x_1,x_2-2)\big],
\end{aligned}
\end{equation}
so the ansatz
\begin{equation}
\psi(x_1,x_2)=A_{12}z_1^{x_1}z_2^{x_2}+A_{21}z_2^{x_1}z_1^{x_2},\qquad x_1<x_2,
\end{equation}
has additive energy $E=\eps(z_1)+\eps(z_2)$. The collision equations obtained from the exact action of $H$ on $|x,x+2\rangle$ and $|x,x+1\rangle$ are
\begin{equation}
E\,\psi(x,x+2)=4\Delta\big[\psi(x-1,x+2)+\psi(x,x+3)\big]+2\ii\,\psi(x,x+4)-2\ii\,\psi(x-2,x+2),
\end{equation}
\begin{equation}
E\,\psi(x,x+1)=2\ii\,\psi(x-1,x)+2\ii\,\psi(x,x+3)-2\ii\,\psi(x-2,x+1)-2\ii\,\psi(x+1,x+2).
\end{equation}
After substituting the two-plane-wave form and the additive energy, the distance-two equation reduces to
\begin{equation}
2\ii(1+z_1z_2)\Big[(1-z_1z_2+2\ii\Delta z_2)A_{12}+(1-z_1z_2+2\ii\Delta z_1)A_{21}\Big]=0,
\end{equation}
and the distance-one equation is the same relation multiplied by $(z_1+z_2)/(z_1z_2)$. Hence, for generic roots with $z_1z_2\ne -1$,
\begin{equation}
\frac{A_{21}}{A_{12}}\equiv S(z_1,z_2)=-\frac{1-z_1z_2+2\ii\Delta z_2}{1-z_1z_2+2\ii\Delta z_1}.
\end{equation}
The exceptional manifold $z_1z_2=-1$ yields a zero-energy degenerate sector not fixed by the scalar $S$-matrix and must
be treated separately \cite{nepo-singular-aba}.

For three magnons, configurations with a single short gap reproduce the two-body collision equation with a spectator plane wave. The genuinely new local intersection patterns are
\begin{equation}
(x,x+1,x+2),\qquad (x,x+1,x+3),\qquad (x,x+2,x+3),\qquad (x,x+2,x+4).
\end{equation}
Projecting $H|\Psi_3\rangle=E|\Psi_3\rangle$ onto these basis states gives
\begin{align}
E\,\psi(x,x+1,x+2)={}&2\ii\,\psi(x-1,x,x+2)-2\ii\,\psi(x-2,x+1,x+2) \notag\\
&-2\ii\,\psi(x,x+2,x+3)+2\ii\,\psi(x,x+1,x+4),
\\
E\,\psi(x,x+1,x+3)={}&2\ii\,\psi(x-1,x,x+3)-2\ii\,\psi(x-2,x+1,x+3) \notag\\
&+4\Delta\big[\psi(x,x+1,x+4)-\psi(x,x+2,x+3)\big] \notag\\
&+2\ii\,\psi(x,x+1,x+5)-2\ii\,\psi(x+1,x+2,x+3),
\\
E\,\psi(x,x+2,x+3)={}&4\Delta\big[\psi(x-1,x+2,x+3)-\psi(x,x+1,x+3)\big] \notag\\
&+2\ii\,\psi(x,x+1,x+2)-2\ii\,\psi(x-2,x+2,x+3) \notag\\
&+2\ii\,\psi(x,x+2,x+5)-2\ii\,\psi(x,x+3,x+4),
\\
E\,\psi(x,x+2,x+4)={}&4\Delta\big[\psi(x-1,x+2,x+4)+\psi(x,x+2,x+5)\big] \notag\\
&+2\ii\,\psi(x,x+2,x+6)-2\ii\,\psi(x-2,x+2,x+4).
\end{align}
Now write
\begin{equation}
\psi(x_1,x_2,x_3)=\sum_{P\in S_3}A(P)\,z_{P_1}^{x_1}z_{P_2}^{x_2}z_{P_3}^{x_3},
\qquad A_{abc}\equiv A(a,b,c),
\end{equation}
and abbreviate $S_{ab}=S(z_a,z_b)$. With $A_{123}=1$, the adjacent exchange rule gives
\begin{equation}
A_{213}=S_{12},\qquad A_{132}=S_{23},\qquad A_{231}=S_{12}S_{13},
\end{equation}
\begin{equation}
A_{312}=S_{13}S_{23},\qquad A_{321}=S_{12}S_{13}S_{23}.
\end{equation}
For the four displayed three-particle equations, we find that the above ansatz is a solution.
Thus the three-particle contact equations impose no new constraint beyond the two-body relation: the scattering is fully factorized and there is no diffraction.

With this consistency check in hand, we make the Ansatz for the generic $N$-magnon wavefunction
\begin{equation}
\psi(x_1,\dots,x_N)=\sum_{P\in S_N}A(P)\prod_{a=1}^N z_{P_a}^{x_a},\qquad 1\le x_1<\cdots<x_N\le L,
\end{equation}
with exchange rule $A(\dots,b,a,\dots)=S(z_a,z_b)A(\dots,a,b,\dots)$. Because $S$ is a scalar, the exchange relations are path independent, and with $A(1,2,\dots,N)=1$ one may take
\begin{equation}
A(P)=\prod_{a<b,\,P_a>P_b}S(z_{P_b},z_{P_a}).
\end{equation}
The energy remains additive,
\begin{equation}
  \label{Y1E}
E=\sum_{j=1}^N\eps(z_j),
\end{equation}
and periodic boundary conditions
\begin{equation}
\psi(x_1,\dots,x_N)=\psi(x_2,\dots,x_N,x_1+L)
\end{equation}
give the Bethe equations
\begin{equation}
  \label{Y1BE}
\e^{\ii Lp_j}=\prod_{k\ne j}\left[-\frac{1-\e^{\ii(p_j+p_k)}+2\ii\Delta\e^{\ii p_j}}{1-\e^{\ii(p_j+p_k)}+2\ii\Delta\e^{\ii p_k}}\right],\qquad E=\sum_{j=1}^N\big(8\Delta\cos p_j-4\sin 2p_j\big),
\end{equation}
where $z_j=\e^{\ii p_j}$. At $\Delta=0$, and away from the exceptional pairwise manifolds $z_jz_k=\pm1$, one recovers
$S=-1$ and the free-fermion quantization condition $\e^{\ii Lp_j}=(-1)^{N-1}$. 

In this work we do not prove the correctness of the $N$-particle Bethe Ansatz based on $H$, because eventually we know
that $H$ is just a higher charge of an established model. However, we confirm the correctness of the
Bethe equations by numerical checks. This is presented in Appendix \ref{app:Y1}.

\section{Solution of model Y2}
\label{sec:Y2}

Now we treat  the periodic spin-$\tfrac12$ chain of even length $L$ with Hamiltonian
\begin{equation}
H=\sum_{j=1}^L \Bigl(
\sigma^+_j \ee^{\ii\gamma\sigma^z_{j+1}}\sigma^-_{j+2}
+\sigma^-_j\ee^{-\ii\gamma\sigma^z_{j+1}}\sigma^+_{j+2}
+\frac{\sin(2\gamma)}{2}\bigl(\sigma^z_j\sigma^z_{j+1}-1\bigr)
\Bigr),
\end{equation}
with periodic identification $j\equiv j+L$ and $\sigma^\pm=(\sigma^x\pm\ii\sigma^y)/2$. We regard the state with all
spins up as the reference state, and the down spins as excitations.

The hopping terms move a
down spin by two sites, so the model conserves not only the total particle number $N=\sum_j n_j$, with
$n_j=(1-\sigma_j^z)/2$, but also the two sublattice  occupations
\begin{equation}
N_{\mathrm{odd}}=\sum_{j\ \mathrm{odd}} n_j,
\qquad
N_{\mathrm{even}}=\sum_{j\ \mathrm{even}} n_j,
\qquad
[H,N_{\mathrm{odd}}]=[H,N_{\mathrm{even}}]=0.
\end{equation}
In order to have the two conservation laws in a finite volume we assume that the length of the chain is even. The full
solution of the chain with an odd length $L$ would be more involved, and we do not treat it in this work.

\subsection{The one particle problem}

In the one-magnon sector a direct action of $H$ on the basis vector $\lvert x\rangle$ with a single down-spin at site $x$ gives
\begin{equation}
H\lvert x\rangle = \ee^{-\ii\gamma}\lvert x-2\rangle + \ee^{\ii\gamma}\lvert x+2\rangle - 2\sin(2\gamma)\lvert x\rangle,
\end{equation}
so a plane wave $\psi(x)=\ee^{\ii kx}$ has exact dispersion
\begin{equation}
  \label{epsY2}
\varepsilon(k)=\ee^{\ii\gamma+2\ii k}+\ee^{-\ii\gamma-2\ii k}-2\sin(2\gamma)=2\cos(2k+\gamma)-2\sin(2\gamma).
\end{equation}

\subsection{The two-particle problem}

A two-magnon basis state is
\[
\ket{x,y}:=\sigma_x^-\sigma_y^-\ket{\vac},
\qquad x<y,
\]
and we write an eigenstate as
\[
\ket{\Psi}=\sum_{x<y}\psi(x,y)\ket{x,y}.
\]
The eigenvalue equation $H\ket{\Psi}=E\ket{\Psi}$ becomes a set of finite-difference
equations for $\psi(x,y)$.

Far from collisions, each magnon behaves as a one-particle excitation hopping by $\pm2$
with a phase that depends on the intermediate site.
If the magnons are sufficiently separated, the intermediate sites are up spins,
$\sigma^z=1$, so the hop phases are constants:
\[
\text{hop right by }+2:\quad \ee^{\ii\gamma\sigma^z_{\,\text{mid}}}=\ee^{+\ii\gamma},
\qquad
\text{hop left by }-2:\quad \ee^{-\ii\gamma\sigma^z_{\,\text{mid}}}=\ee^{-\ii\gamma}.
\]
Also, the Ising term contributes only on bonds where spins differ.
For two well-separated magnons there are $4$ domain walls (four $\uparrow\downarrow$ or
$\downarrow\uparrow$ bonds), each contributing
\[
\frac{\sin(2\gamma)}{2}\big((-1)-1\big)=-\sin(2\gamma),
\]
so the total diagonal Ising contribution is $-4\sin(2\gamma)$.

Hence in the bulk region (no interaction), the Schr\"odinger equation has the form
\begin{align}
E\,\psi(x,y)
&=
-4\sin(2\gamma)\,\psi(x,y)
+\ee^{+\ii\gamma}\,\psi(x+2,y)+\ee^{-\ii\gamma}\,\psi(x-2,y)\nonumber\\
&\hspace{2cm}
+\ee^{+\ii\gamma}\,\psi(x,y+2)+\ee^{-\ii\gamma}\,\psi(x,y-2),
\label{eq:bulk}
\end{align}
where all shifted coordinates remain ordered and allowed.

We now make the coordinate Bethe Ansatz in each ordering region:
\begin{equation}
\psi(x,y)=A_{12}\,\ee^{\ii(k_1 x+k_2 y)}+A_{21}\,\ee^{\ii(k_2 x+k_1 y)}.
\label{eq:CBA2}
\end{equation}
Inserting \eqref{eq:CBA2} into \eqref{eq:bulk} and dividing out $\ee^{\ii(k_1 x+k_2 y)}$
(and similarly for the other term) gives additivity of the energy:
\begin{equation}
E=\varepsilon(k_1)+\varepsilon(k_2),
\qquad
\varepsilon(k)=\ee^{\ii\gamma}\ee^{2\ii k}+\ee^{-\ii\gamma}\ee^{-2\ii k}-2\sin(2\gamma)
=2\cos(2k+\gamma)-2\sin(2\gamma).
\label{eq:addE}
\end{equation}
Up to this point, $A_{12},A_{21}$ are unconstrained: the constraints come from \emph{collision}
equations at minimal separations.

The sub-lattice occupations are separately preserved, and this implies that the Ansatz \eqref{eq:CBA2} is actually too
simple, and we need to take into account the choices of sub-lattice indices. We will derive this step by step in the
following. We use the notations O and E for the odd and even sub-lattices, respectively.

\subsubsection{Case I: same-sublattice collision}

If both magnons live on the same sublattice, the relative distance is even and the minimal
separation is $y=x+2$. At $y=x+2$, hops that would land on the occupied site are forbidden, so:
\begin{itemize}
\item the left magnon at $x$ cannot hop to $x+2$ (occupied),
\item the right magnon at $x+2$ cannot hop to $x$ (occupied),
\item the left magnon can hop to $x-2$ with phase $\ee^{-\ii\gamma}$ (mid-spin up),
\item the right magnon can hop to $x+4$ with phase $\ee^{+\ii\gamma}$ (mid-spin up).
\end{itemize}
Crucially, the Ising diagonal energy at distance $2$ is still $-4\sin(2\gamma)$
(the configuration $\uparrow\downarrow\uparrow\downarrow\uparrow$ has four domain walls, the same as two separated
magnons), 
so the boundary equation at $y=x+2$ is
\begin{equation}
\Big(E+4\sin(2\gamma)\Big)\psi(x,x+2)
=\ee^{-\ii\gamma}\psi(x-2,x+2)+\ee^{+\ii\gamma}\psi(x,x+4).
\label{eq:bdry_same}
\end{equation}

Now substitute the bulk ansatz \eqref{eq:CBA2}. For simplicity set $x=0$ (translation invariance).
Then $\psi(0,2)=A_{12}\ee^{2\ii k_2}+A_{21}\ee^{2\ii k_1}$, etc.
Using the additive energy \eqref{eq:addE}, one finds that \eqref{eq:bdry_same} reduces to
\begin{equation}
\Big(A_{12}+A_{21}\Big)\Big(\ee^{\ii\gamma}\ee^{2\ii(k_1+k_2)}+\ee^{-\ii\gamma}\Big)=0.
\end{equation}
For generic $(k_1,k_2,\gamma)$ the second factor is nonzero, hence
\begin{equation}
A_{21}=-A_{12}
\quad\Longrightarrow\quad
S_{\OO\OO}(k_1,k_2)=S_{\EE\EE}(k_1,k_2)=-1.
\end{equation}
Thus in the same-sublattice sectors the two magnons scatter by a momentum-independent fermionic sign.

\subsubsection{Case II: mixed-sublattice collision and non-diagonal scattering}

Now consider one magnon on an odd site and one on an even site. The relative distance is odd,
and the minimal separation is \emph{adjacency} $y=x+1$.
At adjacency, a magnon can hop by $2$ \emph{over} the other magnon, swapping their order along the chain.
This produces non-diagonal scattering.

In the ordered region $x<y$, there are two possible parity orderings:
\[
(\OO\EE):\ x\ \text{odd},\ y\ \text{even},
\qquad
(\EE\OO):\ x\ \text{even},\ y\ \text{odd}.
\]
We therefore define two components:
\[
\psi_{\OO\EE}(x,y)\quad (x\text{ odd},y\text{ even}),\qquad
\psi_{\EE\OO}(x,y)\quad (x\text{ even},y\text{ odd}),
\]
and in the bulk (distance $\ge3$) we use the two-plane-wave ansatz for each component:
\begin{align}
\psi_{\OO\EE}(x,y)&=A_{12}\,\ee^{\ii(k_1 x+k_2 y)}+A_{21}\,\ee^{\ii(k_2 x+k_1 y)},\\
\psi_{\EE\OO}(x,y)&=B_{12}\,\ee^{\ii(k_1 x+k_2 y)}+B_{21}\,\ee^{\ii(k_2 x+k_1 y)}.
\end{align}
The energy is still $E=\varepsilon(k_1)+\varepsilon(k_2)$ from the bulk equation.

Take the adjacent configuration $(\OO\EE)$ at sites $(x,x+1)$ with $x$ odd.
There are \emph{four} allowed hops (none land on an occupied site):
\begin{itemize}
\item left magnon hops left: $(x,x+1)\to(x-2,x+1)$ with phase $\ee^{-\ii\gamma}$ (mid-spin up),
\item right magnon hops right: $(x,x+1)\to(x,x+3)$ with phase $\ee^{+\ii\gamma}$ (mid-spin up),
\item left magnon hops right over the right magnon: $(x,x+1)\to(x+1,x+2)$ (now \emph{order flips} to $\EE\OO$)
with phase $\ee^{-\ii\gamma}$ because the intermediate spin at $x+1$ is \emph{down},
\item right magnon hops left over the left magnon: $(x,x+1)\to(x-1,x)$ (order flips to $\EE\OO$)
with phase $\ee^{+\ii\gamma}$ because the intermediate spin at $x$ is \emph{down}.
\end{itemize}

\paragraph{Diagonal Ising term at adjacency.}
In the adjacent configuration $\downarrow\downarrow$ there is \emph{no} domain wall between the two magnons,
so compared to two separated magnons there are only $2$ domain walls (instead of $4$).
Therefore the Ising diagonal energy at adjacency is $-2\sin(2\gamma)$ (not $-4\sin(2\gamma)$).

Putting everything together, the Schr\"odinger equation at adjacency $(\OO\EE)$ is
\begin{align}
\Big(E+2\sin(2\gamma)\Big)\psi_{\OO\EE}(x,x+1)
&=\ee^{-\ii\gamma}\psi_{\OO\EE}(x-2,x+1)+\ee^{+\ii\gamma}\psi_{\OO\EE}(x,x+3)\nonumber\\
&\quad+\ee^{-\ii\gamma}\psi_{\EE\OO}(x+1,x+2)+\ee^{+\ii\gamma}\psi_{\EE\OO}(x-1,x).
\label{eq:adj_OE}
\end{align}
A completely analogous equation holds for adjacency $(\EE\OO)$ (take $x$ even and repeat the list of hops).

It is convenient to introduce the compact variables
\begin{equation}
q:=\ee^{\ii\gamma},\qquad z_a:=\ee^{\ii k_a},\qquad y_a:=z_a^2=\ee^{2\ii k_a}\qquad (a=1,2).
\end{equation}
Because of translation invariance, we may impose the adjacency equations at the two canonical adjacent points
\[
(\OO\EE):\ (x,y)=(1,2),\qquad (\EE\OO):\ (x,y)=(0,1),
\]
and substitute the plane-wave forms for all amplitudes appearing in \eqref{eq:adj_OE} and its partner.
Using $E=\varepsilon(k_1)+\varepsilon(k_2)$, the explicit dependence on $x$ cancels and we obtain
\emph{two linear equations} for the two unknown outgoing amplitudes $(A_{21},B_{21})$ in terms of the incoming
ones $(A_{12},B_{12})$.

The result can be written as a $2\times2$ scattering matrix acting on the internal ordering space:
\begin{equation}
\begin{pmatrix}A_{21}\\[1mm]B_{21}\end{pmatrix}
=
\check S(k_1,k_2)\,
\begin{pmatrix}A_{12}\\[1mm]B_{12}\end{pmatrix},
\qquad
\check S(k_1,k_2)=
\begin{pmatrix}
b(k_1,k_2) & c(k_1,k_2)\\
c(k_1,k_2) & b(k_1,k_2)
\end{pmatrix}.
\label{eq:checkR_mixed}
\end{equation}
Solving the $2\times2$ linear system and simplifying gives the compact closed forms
\begin{equation}
b(k_1,k_2)=
\frac{z_1z_2\,(q^4-1)\,(\ii q+y_1)(\ii q+y_2)}{D(y_1,y_2)},
\qquad
c(k_1,k_2)=
\frac{q^2\,(q^2+y_1y_2)\,(y_1-y_2)}{D(y_1,y_2)},
\label{eq:bc_final}
\end{equation}
with the denominator polynomial
\begin{equation}
D(y_1,y_2)=
q^6 y_1-2\ii q^5 y_1y_2-q^4y_1y_2^2-q^2 y_2+2\ii q y_1y_2+y_1^2 y_2.
\label{eq:D_final}
\end{equation}

The result satisfies the following consistency checks:
\begin{itemize}
\item \textbf{Regularity:} set $k_1=k_2\Rightarrow y_1=y_2$. Then \eqref{eq:bc_final} gives
$c(k,k)=0$, while a direct substitution shows $b(k,k)=-1$. Hence
\[
\check S(k,k)=-1
\]
\item \textbf{Exchange symmetry:} It can be checked directly that
  \begin{equation}
    \check S(k_1,k_2) \check S(k_2,k_1)=1
  \end{equation}  
\end{itemize}

Combining the same-sublattice result $S=-1$ with the mixed block \eqref{eq:checkR_mixed},
the full braid scattering matrix in the ordered parity basis
$(\OO\OO,\OO\EE,\EE\OO,\EE\EE)$ is
\begin{equation}
\check R(k_1,k_2)=
\begin{pmatrix}
-1 & 0 & 0 & 0\\
0 & b(k_1,k_2) & c(k_1,k_2) & 0\\
0 & c(k_1,k_2) & b(k_1,k_2) & 0\\
0 & 0 & 0 & -1
\end{pmatrix}.
\end{equation}
This is the complete two-particle solution needed for the (nested) Bethe Ansatz in mixed sectors.

\subsection{Properties of the $R$-matrix}

Now we show that the $R$-matrix coincides with that of the 6-vertex model.

Let us define $Q=q^2$. We introduce the reparametrized rapidity
\[
\rho(k):=e^{ik}+\mathrm i qe^{-ik},
\qquad
u_{12}:=\frac{\rho(k_1)}{\rho(k_2)}.
\]
Equivalently, an additive rapidity can be found as
\[
\theta(k):=\log \rho(k),
\qquad
\theta_{12}:=\theta(k_1)-\theta(k_2),
\qquad
u_{12}=e^{\theta_{12}}.
\]

Now the denominator factorizes as
\[
D(y_1,y_2)
=
y_2(y_1+\mathrm i q)^2-q^4 y_1(y_2+\mathrm i q)^2
=
z_1^2z_2^2\bigl(\rho_1^2-Q^2\rho_2^2\bigr),
\]
where \(\rho_a:=\rho(k_a)\). Moreover,
\[
z_1z_2(q^4-1)(\mathrm i q+y_1)(\mathrm i q+y_2)
=
(Q^2-1)z_1^2z_2^2\,\rho_1\rho_2,
\]
and
\[
q^2(q^2+y_1y_2)(y_1-y_2)
=
Qz_1^2z_2^2(\rho_1^2-\rho_2^2).
\]
Hence
\[
b(k_1,k_2)=\frac{(Q^2-1)u_{12}}{u_{12}^2-Q^2},
\qquad
c(k_1,k_2)=\frac{Q(u_{12}^2-1)}{u_{12}^2-Q^2}.
\]

Therefore the checked matrix becomes
\[
-\check R(k_1,k_2)
=
\frac{1}{Q u_{12}^{-1}-Q^{-1}u_{12}}
\begin{pmatrix}
Q u_{12}^{-1}-Q^{-1}u_{12} & 0 & 0 & 0\\
0 & Q-Q^{-1} & u_{12}-u_{12}^{-1} & 0\\
0 & u_{12}-u_{12}^{-1} & Q-Q^{-1} & 0\\
0 & 0 & 0 & Q u_{12}^{-1}-Q^{-1}u_{12}
\end{pmatrix}.
\]

This is the standard checked trigonometric six-vertex matrix in regular
normalization, and one has the exact identification
\begin{equation}
{
\check R(k_1,k_2)
=
-\check R_{6\mathrm v}^{\mathrm{reg}}
\!\left(
\frac{\rho(k_1)}{\rho(k_2)}\,;\,q^2
\right)
}.
\end{equation}

If we undo the check, i.e. pass to \(R=P\check R\), and 
change to additive rapidities,
\[
u_{12}=e^{\theta_{12}},
\qquad
Q=e^{\eta},
\]
we get
\begin{equation}
-R(\theta_{12})
=
\frac{1}{\sinh(\eta-\theta_{12})}
\begin{pmatrix}
\sinh(\eta-\theta_{12}) & 0 & 0 & 0\\
0 & \sinh\theta_{12} & \sinh\eta & 0\\
0 & \sinh\eta & \sinh\theta_{12} & 0\\
0 & 0 & 0 & \sinh(\eta-\theta_{12})
\end{pmatrix}.
\end{equation}
Equivalently,
\[
a(\theta)=1,
\qquad
b(\theta)=\frac{\sinh\theta}{\sinh(\eta-\theta)},
\qquad
c(\theta)=\frac{\sinh\eta}{\sinh(\eta-\theta)}.
\]

So in the usual trigonometric six-vertex notation one may write
\[
{
-R_{\mathrm{full}}(\theta)
=
R_{6\mathrm v}^{\mathrm{reg}}(\theta;\eta)
},
\qquad
\eta=2\log q.
\]

Finally, the anisotropy parameter is
\[
\Delta=\frac{Q+Q^{-1}}{2}
=\frac{q^2+q^{-2}}{2}.
\]

\subsection{General $N$-magnon coordinate Bethe Ansatz with an internal two-state space}

Each magnon carries an internal label $a\in\{\mathrm{O},\mathrm{E}\}$ telling on which sublattice
(odd/even) it lives.  The two-body analysis shows that when two magnons are exchanged,
their amplitudes are mixed by the $4\times 4$ braid matrix $\check R$; in particular the mixed
$\mathrm{OE}/\mathrm{EO}$ channel has a non-diagonal $2\times2$ block.

For $N$ magnons with ordered coordinates
\[
x_1<x_2<\cdots<x_N,
\]
the coordinate Bethe Ansatz in that region is written as
\begin{equation}
\Psi_{a_1\ldots a_N}(x_1,\ldots,x_N)
=
\sum_{P\in S_N}
A_{a_1\ldots a_N}(P)\,
\exp\!\Big(\mathrm{i}\sum_{j=1}^N k_{Pj}x_j\Big),
\label{eq:CBA_N}
\end{equation}
where:
\begin{itemize}
\item $P$ is a permutation of the momenta $(k_1,\ldots,k_N)$,
\item $A(P)$ is a vector in the internal space $V^{\otimes N}$ with $V=\mathrm{span}\{\mathrm{O},\mathrm{E}\}$,
\item the energy is additive in the bulk: $E=\sum_{j=1}^N \varepsilon(k_j)$.
\end{itemize}

As usual, the bulk difference equation (valid when no particles are at the minimal allowed separation)
fixes the dispersion $\varepsilon(k)$ and implies additivity.  All nontrivial constraints on the
coefficients $A(P)$ come from \emph{boundary/contact hyperplanes} where two neighboring coordinates
approach the minimal allowed distance.

Let us now discuss our claim that this Hamiltonian has no ``genuine'' three-body contact term. In other words, the
three-particle Bethe Ansatz gives correct eigenfunctions.

Although the hopping term is written on three sites,
\[
\sigma_j^{+}\, \mathrm{e}^{\mathrm{i}\gamma\sigma^z_{j+1}}\,\sigma_{j+2}^{-}
\ +\ 
\sigma_j^{-}\, \mathrm{e}^{-\mathrm{i}\gamma\sigma^z_{j+1}}\,\sigma_{j+2}^{+},
\]
it moves \emph{one} magnon between $j$ and $j+2$, and the only effect of the middle site $j+1$ is to
multiply this \emph{two-body hop} by a phase depending on whether $j+1$ is occupied or not.
Because of the hard-core constraint (each site is either up or down), the phase is influenced by at most
\emph{one} magnon.  There is no term in $H$ that changes the positions of three magnons simultaneously.
The remaining interaction $\frac{\sin(2\gamma)}{2}(\sigma^z_j\sigma^z_{j+1}-1)$ is purely two-body and
diagonal.

Therefore, in any short-distance cluster of three magnons, every nontrivial process is a sequence of
pairwise processes (two magnons exchanging or one magnon hopping past one neighbor).  In the coordinate
Bethe Ansatz, this means that every short-distance constraint can be built from the same two-body
matching rule.

\subsection{The nested Bethe Ansatz}

Now we compute the solution of the nested problem. We follow standard derivations \cite{fedor-lectures}, therefore we
keep the details minimal.

To write the nested equations explicitly it is convenient to pass to the vertex matrix $R=P\check R$ and then remove the constant diagonal weight by setting $\widetilde R=-R$. The normalized mixed weights are therefore $\widetilde a(u,v)=1$, $\widetilde b(u,v)=-c(u,v)$, and $\widetilde c(u,v)=-b(u,v)$ in standard six-vertex notation. For an ordered set of particle momenta $k_1,\dots,k_N$, the normalized monodromy matrix $\widetilde T_a(u)=\widetilde R_{aN}(u,k_N)\cdots \widetilde R_{a1}(u,k_1)$ and transfer matrix $\widetilde t(u)=\mathrm{tr}_a\,\widetilde T_a(u)$ have the algebraic-Bethe-ansatz eigenvalue
\begin{equation}
\widetilde\Lambda(u)=
\prod_{\alpha=1}^M \frac{1}{-c(\lambda_\alpha,u)}
+
\Biggl[
\prod_{j=1}^N \bigl(-c(u,k_j)\bigr)
\Biggr]
\prod_{\alpha=1}^M \frac{1}{-c(u,\lambda_\alpha)},
\end{equation}
where $M=N_{\mathrm{even}}$ if the odd sublattice is used as pseudovacuum. Cancellation of the unwanted terms gives the auxiliary equations
\begin{equation}
  \label{Y2be1}
(-1)^N \prod_{j=1}^N c(\lambda_\alpha,k_j)=
\prod_{\substack{\beta=1\\ \beta\neq\alpha}}^M
\frac{c(\lambda_\alpha,\lambda_\beta)}{c(\lambda_\beta,\lambda_\alpha)},
\qquad
\alpha=1,\dots,M,
\end{equation}
and periodicity of the coordinate wave function gives
\begin{equation}
  \label{Y2be2}
\ee^{\ii k_jL}=
(-1)^{N+M+1}
\prod_{\alpha=1}^M c(\lambda_\alpha,k_j),
\qquad
j=1,\dots,N.
\end{equation}

Once a solution is found to these equations, the energy is computed as
\begin{equation}
  \label{Y2E}
  E=\sum_{j=1}^N \varepsilon(k_j),
\end{equation}
where $\varepsilon(k)$ is given by \eqref{epsY2}.

In the special case of $M=0$ this reduces immediately to the free-fermion equation $\ee^{\ii k_jL}=(-1)^{N-1}$ found
above.

\section{Solution of model Y3 -- the $S$-matrix}

\label{sec:Y3}

We consider the periodic spin-$\tfrac12$ chain of length $L\ge 4$ with Hilbert space $(\C^2)^{\otimes L}$ and Hamiltonian
\begin{equation}
H=\sum_{j=1}^{L} h(j),
\qquad
h(j)=\Pe_{j,j+3}\Pe_{j+1,j+2}-\Pe_{j,j+3}-\Pe_{j+1,j+2}-\Pe_{j,j+2}+2
\end{equation}
where $\Pe_{i,j}$ exchanges the spins at sites $i$ and $j$ and all site labels are understood modulo $L$. In the
$N$-magnon sector we use the ordered basis $\ket{x_1,\dots,x_N}$ with $0\le x_1<\cdots<x_N\le L-1$.

For one magnon, writing $\ket{x}$ for the state with a single down spin at $x$, a direct evaluation gives
\begin{equation}
H\ket{x}=2\ket{x}-\ket{x-2}-\ket{x+2}.
\end{equation}
Hence the plane wave $\psi(x)=z^x$ is an eigenfunction with dispersion
\begin{equation}
\varepsilon(z)=2-z^2-z^{-2},
\qquad
\varepsilon(e^{\ii p})=4\sin^2 p.
\end{equation}
Because $\varepsilon(z)=\varepsilon(-z)$, each one-particle rapidity carries a two-dimensional internal degeneracy, and
this is the origin of the nested coordinate Bethe Ansatz. The transformation $z\to -z$ corresponds to the shift $p\to
p+\pi$ in the lattice momentum.

We could take linear combinations of waves propagating with $z=e^{\ii p}$ and $-z=-e^{\ii p}$, in order to obtain waves
confined to the even and odd sub-lattices. However, the present choice will prove to be useful for the Bethe Ansatz
calculations below.

Similar to the case of the model Y2, here we also assume that the length $L$ of the chain is an even number.

\subsection{The two-particle problem}

For two magnons it is convenient to solve the local scattering problem on the infinite line with ordered coordinates $x<y$ and to impose periodicity only afterwards. If $r=y-x\ge 3$, the two magnons are separated and one finds
\begin{equation}
H\ket{x,y}=4\ket{x,y}-\ket{x-2,y}-\ket{x+2,y}-\ket{x,y-2}-\ket{x,y+2}.
\end{equation}
The only contact relations occur at separations $r=2$ and $r=1$, for which direct evaluation gives
\begin{align}
H\ket{x,x+2}
&=4\ket{x,x+2}-\ket{x-2,x+2}-\ket{x,x+4}-\ket{x-1,x}-\ket{x,x+1} \notag \\
&\quad -\ket{x+1,x+2}-\ket{x+2,x+3}+\ket{x-1,x+1}+\ket{x+1,x+3}, \\
H\ket{x,x+1}
&=6\ket{x,x+1}-\ket{x-2,x}-\ket{x-2,x+1}-\ket{x-1,x}-\ket{x-1,x+1} \notag \\
&\quad -\ket{x,x+2}-\ket{x,x+3}-\ket{x+1,x+2}-\ket{x+1,x+3} \notag \\
&\quad +\ket{x-2,x-1}+\ket{x+2,x+3}.
\end{align}
The coordinate Bethe Ansatz must therefore include the internal signs associated with the one-particle degeneracy. With $\sigma_1,\sigma_2\in\signs$ we write
\begin{equation}
\Psi(x,y)=\sum_{\sigma_1,\sigma_2\in\signs}
\Bigl[A_{12}^{\sigma_1\sigma_2}(\sigma_1 z)^x(\sigma_2 w)^y
+A_{21}^{\sigma_1\sigma_2}(\sigma_1 w)^x(\sigma_2 z)^y\Bigr],
\qquad x<y.
\end{equation}
For $r\ge 3$ the Schr\"odinger equation gives the additive energy $E=\varepsilon(z)+\varepsilon(w)$. Solving the two contact equations at $r=2$ and $r=1$ yields the exchange relation $A_{21}=\check R(z,w)A_{12}$, where in the ordered basis $(\ket{++},\ket{+-},\ket{-+},\ket{--})$ the braided scattering matrix is
\begin{equation}
\check R(z,w)=
\begin{pmatrix}
 a_+(z,w) & 0 & 0 & d(z,w) \\
 0 & b(z,w) & c_+(z,w) & 0 \\
 0 & c_-(z,w) & b(z,w) & 0 \\
 d(z,w) & 0 & 0 & a_-(z,w)
\end{pmatrix}.
\end{equation}
If one introduces
\begin{align}
\Delta(z,w)&=z^2w^2-3z^2+w^2+1, \\
A_{\pm}(z,w)&=z^2w^2+z^2+w^2+1-4zw\pm 2(z^2w-zw^2+z-w), \\
C_{\pm}(z,w)&=z^2w^2+z^2+w^2+1+4zw\pm 2(z^2w+zw^2-z-w),
\end{align}
then the nonzero matrix elements are
\begin{align}
a_{\pm}(z,w)&=-\frac{(z+w)A_{\pm}(z,w)}{2z\,\Delta(z,w)},
&
 b(z,w)&=-\frac{(z^2-1)(w^2-1)(z+w)}{2z\,\Delta(z,w)}, \\
c_{\pm}(z,w)&=-\frac{(z-w)C_{\pm}(z,w)}{2z\,\Delta(z,w)},
&
 d(z,w)&=-\frac{(z^2-1)(w^2-1)(z-w)}{2z\,\Delta(z,w)}.
\end{align}
The matrix $\check R(z,w)$ preserves the sign parity $\sigma_1\sigma_2$, with even block
$\bigl(\begin{smallmatrix}a_+&d\\ d&a_-\end{smallmatrix}\bigr)$ on $\mathrm{span}\{\ket{++},\ket{--}\}$ and odd block
$\bigl(\begin{smallmatrix}b&c_+\\ c_-&b\end{smallmatrix}\bigr)$ on $\mathrm{span}\{\ket{+-},\ket{-+}\}$. 

We have the relations
\begin{align}
  \check R(z,z)&=-1, \\
\check R(z,w)\check R(w,z)&=1
\end{align}
and
\begin{align}
\check R_{23}(z_1,z_2)\check R_{12}(z_1,z_3)\check R_{23}(z_2,z_3)
=
\check R_{12}(z_2,z_3)\check R_{23}(z_1,z_3)\check R_{12}(z_1,z_2).
\end{align}
The un-braided $R$-matrix satisfies
\begin{align}
  R_{12}(z_1,z_2)R_{13}(z_1,z_3)R_{23}(z_2,z_3)
&=R_{23}(z_2,z_3)R_{13}(z_1,z_3)R_{12}(z_1,z_2).
\end{align}

The eigenvalues of the matrix $R(u,z)$ are
\begin{equation}
  \label{Reigs}
  \left\{\rho(u,z),\pm \frac{z}{u}\right\},
\end{equation}
and the eigenvalue $\rho(u,z)$ appears with multiplicity 2.

The weights obey the free-fermion relation \cite{fan-wu}
\begin{equation}
  \label{eq:free-fermion-identity}
a_+(z,w)a_-(z,w)+c_+(z,w)c_-(z,w)=b(z,w)^2+d(z,w)^2,
\end{equation}
which is equivalent to the equality of the determinants of the even and odd $2\times2$ blocks. Thus $\check R(z,w)$ is a
parity-preserving matchgate.

For completeness we derive the first derivative of the matrix $\check R(z,w)$ at the so-called regular points $z=w$. If
we were to define a spin chain problem using this $R$-matrix, then the derivative would give the Hamiltonian density.
We get
\begin{equation}
  \begin{split}
h(w)=& \check R(w,w)^{-1} \left.\partial_z \check R(z,w)\right|_{z=w}=\\
&-\frac{w^4-10w^2+1}{2w(w^2-1)^2}
+\frac{w^2+1}{(w^2-1)^2}(Z_1+Z_2)
+\frac{(w^2+1)^2}{2w(w^2-1)^2}X_1X_2
+\frac{2w}{(w^2-1)^2}Y_1Y_2\\
&-\frac{\mathrm i}{w^2-1}(X_1Y_2-Y_1X_2).  
      \end{split}
    \end{equation}
This is a free fermionic Hamiltonian density. The fact that it has an explicit $w$ dependence corresponds to the
$R$-matrix having a non-difference form.

It can be shown that this $R$-matrix is a specific case of the 8-vertex type free fermionic $R$-matrices known in the
literature. In particular, it is a member of the 8vB family treated in \cite{marius-classification-6}. However, we do
not treat the details of the identification here.

\subsection{The $N$-body problem}

For $N$ magnons with ordered coordinates $x_1<\cdots<x_N$ the coordinate Bethe Ansatz takes the form
\begin{equation}
\Psi(x_1,\dots,x_N)=\sum_{\pi\in S_N}\;\sum_{\sigma_1,\dots,\sigma_N\in\signs}
A_{\pi}^{\sigma_1\dots\sigma_N}\prod_{j=1}^N (\sigma_j z_{\pi(j)})^{x_j},
\end{equation}
where $A_\pi\in(\C^2)^{\otimes N}$ is the internal amplitude vector attached to the permutation $\pi$. Adjacent exchanges are controlled by the same two-body matrix,
\begin{equation}
A_{\pi s_j}=\check R_{j,j+1}(z_{\pi(j)},z_{\pi(j+1)})A_\pi,
\qquad j=1,\dots,N-1,
\end{equation}
where $s_j$ is the adjacent transposition exchanging $j$ and $j+ 1$.
The energy remains additive,
\begin{equation}
E=\sum_{k=1}^N \varepsilon(z_k).
\end{equation}
To formulate the periodicity conditions it is convenient to use the monodromy matrix
\begin{equation}
T_a(u\mid z_1,\dots,z_N)=R_{aN}(u,z_N)\cdots R_{a1}(u,z_1),
\qquad
 t(u)=\tr_a T_a(u\mid z_1,\dots,z_N),
\end{equation}
which acts on the internal space $(\C^2)^{\otimes N}$. The Yang-Baxter equation implies $[t(u),t(v)]=0$ for all $u,v$, so the nested problem is reduced to a commuting family of $2^N\times2^N$ matrices. Using regularity, one obtains the specialized operators
\begin{equation}
\mathcal T_k(\{z\})=-t(z_k)
=R_{k,k-1}(z_k,z_{k-1})\cdots R_{k1}(z_k,z_1)
 R_{kN}(z_k,z_N)\cdots R_{k,k+1}(z_k,z_{k+1}),
\end{equation}
with the convention that an empty product equals $\id$. Periodic boundary conditions are then
\begin{equation}
z_k^L\,\mathcal T_k(\{z\})A=A,
\qquad k=1,\dots,N,
\end{equation}
for a common internal vector $A\in(\C^2)^{\otimes N}$. Equivalently, if $A$ is a simultaneous eigenvector of the commuting family $\{\mathcal T_k\}_{k=1}^N$ with eigenvalues $\Lambda_k(\{z\})$, the nested Bethe equations are
\begin{equation}
  \label{nestedgeneral}
z_k^L\,\Lambda_k(\{z\})=1,
\qquad k=1,\dots,N.
\end{equation}
We stress that with these conventions $\Lambda_k(\{z\})$ is the eigenvalue of the monodromy-type operators $\mathcal
T_k(\{z\})$, and in particular they are eigenvalues of $-t(z_k)$.

The free fermionic nature of the $R$-matrix implies that the transfer matrix can be diagonalized using free
fermions. We will perform this computation later in Section \ref{sec:freefermions}. However, first we discuss two simple
branches of eigenvectors of the transfer matrix.

\subsection{Explicit product branches}

Two distinguished branches of the nested problem can be written explicitly because the local two-body problem has two
factorized 
eigenvectors. Define 
\begin{equation}
\alpha_\tau(z)=\begin{pmatrix} z+1 \\ \tau(z-1) \end{pmatrix}.
\end{equation}
Here $\tau=\pm$, and the two signs correspond to two different factorized eigenvectors.

A direct substitution gives
\begin{equation}
R(z,w)\bigl(\alpha_\tau(z)\otimes \alpha_\tau(w)\bigr)=\rho(z,w)\,\alpha_\tau(z)\otimes \alpha_\tau(w),
\end{equation}
where
\begin{equation}
  \rho(z,w)=-\frac{\Delta(w,z)}{\Delta(z,w)}.
\end{equation}

Consider now the  tensor product of these one-site vectors:
\begin{equation}
A_\tau(z_1,\dots,z_N)=\alpha_\tau(z_1)\otimes\cdots\otimes \alpha_\tau(z_N)
\end{equation}
We will see that the symmetric and anti-symmetric combinations of the vectors $A_\pm$ (corresponding to their parity
projections)   are simple eigenstates of 
the transfer matrix. This way we can obtain two branches of the full nested problem, leading to a single set of Bethe equations.

The computations are more transparent after a rapidity-dependent change of basis.
Define the local $2\times2$ matrix
\begin{equation}
G(z)=\begin{pmatrix} z+1 & z+1 \\ z-1 & 1-z \end{pmatrix}
=\bigl(\alpha_+(z),\alpha_-(z)\bigr),
\end{equation}
and for every permutation $\pi\in S_N$ introduce the dressed amplitudes
\begin{equation}
\widetilde A_\pi=
\bigl(G(z_{\pi(1)})^{-1}\otimes\cdots\otimes G(z_{\pi(N)})^{-1}\bigr)A_\pi.
\end{equation}
Because the basis depends on the rapidity carried by each tensor factor, the exchange relation becomes
\begin{equation}
\widetilde A_{\pi s_j}=
\widetilde{\check R}_{j,j+1}(z_{\pi(j)},z_{\pi(j+1)})\,\widetilde A_\pi,
\end{equation}
with the gauged braided matrix
\begin{equation}
\widetilde{\check R}(u,z)=
\bigl(G(z)^{-1}\otimes G(u)^{-1}\bigr)
\check R(u,z)
\bigl(G(u)\otimes G(z)\bigr)
=
\begin{pmatrix}
\rho(u,z) & \beta(u,z) & \gamma(u,z) & 0 \\
0 & -\dfrac{z}{u} & 0 & 0 \\
0 & 0 & -\dfrac{z}{u} & 0 \\
0 & \gamma(u,z) & \beta(u,z) & \rho(u,z)
\end{pmatrix},
\label{eq:gaugedRfreefermion}
\end{equation}
where
\begin{equation}
\rho(u,z)=-\frac{\Delta(z,u)}{\Delta(u,z)},
\qquad
\beta(u,z)=\frac{2z(z^2-u^2)}{\Delta(u,z)},
\qquad
\gamma(u,z)=\frac{2(z^2-u^2)}{u\,\Delta(u,z)}.
\end{equation}
The key point is that \eqref{eq:gaugedRfreefermion} is the standard triangular free-fermion local weight: the two-body
scattering in the nested problem is no longer interacting. 

Interestingly, the basis transformation yields
\begin{equation}
  G^{-1}(u)ZG(u)=X,
\end{equation}
which implies that the fermionic parity operator will take the form
\begin{equation}
  \tilde Q=\prod_{j=1}^N X_j
\end{equation}
in the new basis.

Let
\begin{equation}
\mathcal{G}(\{z\})=G_1(z_1)\otimes\cdots\otimes G_N(z_N)
\end{equation}
act on the physical spaces. Consider the monodromy matrix in the new basis
\begin{equation}
\widetilde T_a(u)=\widetilde R_{aN}(u,z_N)\cdots \widetilde R_{a1}(u,z_1)
=
G_a(u)^{-1}\mathcal{G}(\{z\})^{-1}T_a(u)\mathcal{G}(\{z\})G_a(u).
\end{equation}
After taking the auxiliary trace
\begin{equation}
\widetilde t(u)=\mathcal{G}(\{z\})^{-1}t(u)\mathcal{G}(\{z\}).
\label{eq:gauged-transfer-similarity}
\end{equation}
So the transfer matrix built from $\widetilde R$ is similar to the original
one.

Now define
\begin{equation}
\widetilde A_+=e_+^{\otimes N},
\qquad
\widetilde A_-=e_-^{\otimes N},
\end{equation}
where $e_\pm$ are the basis vectors in the new basis.

By construction,
\begin{equation}
\mathcal{G}(\{z\})\widetilde A_\tau
=
\alpha_\tau(z_1)\otimes\cdots\otimes\alpha_\tau(z_N)
=:A_\tau(z_1,\dots,z_N),
\qquad \tau=\pm.
\end{equation}
Moreover,
\begin{equation}
\widetilde T_a(u)\bigl(e_{\tau,a}\otimes\widetilde A_\tau\bigr)
=
\Bigl[\prod_{j=1}^N \rho(u,z_j)\Bigr]
\bigl(e_{\tau,a}\otimes\widetilde A_\tau\bigr).
\end{equation}

Equation \eqref{eq:gaugedRfreefermion} immediately implies that the two-dimensional subspace
\begin{equation}
\mathcal V_0=\mathrm{span}\{A_+,A_-\}
\end{equation}
is invariant under the full transfer matrix $t(u)$. Indeed, starting from $A_+$ there are only two paths that remain
inside $\mathcal V_0$ after one full auxiliary sweep: 
\begin{itemize}
\item the auxiliary state starts with $+$ and every local vertex contributes the $\rho$-weight, producing again $A_+$,
\item the auxiliary state starts with $-$ and every local vertex contributes the $\beta$-weight, producing $A_-$.
\end{itemize}
The same reasoning applies to $A_-$. 

This implies
\begin{align}
\widetilde t(u)\widetilde A_+
&=
\Bigl[\prod_{j=1}^N \rho(u,z_j)\Bigr]\widetilde A_+
+
\Bigl[\prod_{j=1}^N \beta(u,z_j)\Bigr]\widetilde A_-,
\\
\widetilde t(u)\widetilde A_-
&=
\Bigl[\prod_{j=1}^N \beta(u,z_j)\Bigr]\widetilde A_+
+
\Bigl[\prod_{j=1}^N \rho(u,z_j)\Bigr]\widetilde A_-.
\end{align}
Thus for generic $u$ the true vacuum eigenvectors are
$\widetilde A_+\pm \widetilde A_-$, while the simple product states
$\widetilde A_\pm$ themselves become eigenvectors at the special points
$u=z_k$. Indeed,
\begin{equation}
\beta(z_k,z_k)=0,
\qquad
\rho(z_k,z_k)=-1,
\end{equation}
so
\begin{equation}
\widetilde t(z_k)\widetilde A_\tau
=
-\prod_{\substack{j=1\\ j\neq k}}^N \rho(z_k,z_j)\,\widetilde A_\tau.
\end{equation}
Conjugating back with \eqref{eq:gauged-transfer-similarity} gives
\begin{equation}
  \label{tmprod}
t(z_k)A_\tau(z_1,\dots,z_N)
=
-\prod_{\substack{j=1\\ j\neq k}}^N \rho(z_k,z_j)\,A_\tau(z_1,\dots,z_N).
\end{equation}

Hence the genuine transfer-matrix eigenvectors in this vacuum sector are the symmetric and antisymmetric combinations
\begin{equation}
\Omega_\pm=A_+\pm A_-,
\end{equation}
The action of the fermionic parity operator yields
\begin{equation}
Q\Omega_\pm=\pm \Omega_\pm.
\end{equation}
This way we confirmed that these special eigenstates of the transfer matrix are also eigenstates of the fermionic parity.

Substituting \eqref{tmprod} into \eqref{nestedgeneral} we get the single set of Bethe equations
\begin{equation}
z_k^L \prod_{\substack{j=1\\ j\neq k}}^N \rho(z_k,z_j)=1,
\qquad k=1,\dots,N,
\end{equation}
which can also be written in the explicit rational form
\begin{equation}
(-1)^{N-1}
\prod_{\substack{j=1 \\ j\ne k}}^{N}
\frac{z_k^2z_j^2+z_k^2-3z_j^2+1}{z_k^2z_j^2-3z_k^2+z_j^2+1}
=
 z_k^{-L},
\qquad k=1,\dots,N.
\label{eq:productbranchBAErational}
\end{equation}

Interestingly, there is an additive spectral parameter for this branch.
Introduce
\[
x(z):=\frac{1+z^2}{1-z^2}.
\]
Then
\[
\Delta(u,z)=u^2z^2-3u^2+z^2+1
=(1-u^2)(1-z^2)\bigl(1+x(z)-x(u)\bigr),
\]
and therefore
\[
\rho(u,z)=-\frac{\Delta(z,u)}{\Delta(u,z)}
=\frac{x(u)-x(z)+1}{x(u)-x(z)-1}.
\]
Thus \(x(z)\) is an additive spectral variable for these special branches.

At the same time, it appears that there is no global
additive spectral parameter. This can be seen already from the eigenvalues of the $R$-matrix \eqref{Reigs}: the other
eigenvalue is not additive when expressed via the $x$-parametrization.

\section{Solution of model Y3 -- free fermions at the nested level}

\label{sec:freefermions}

In this Section we compute the eigenvalues of the transfer matrix of model Y3. These eigenvalues have to be substituted
into equations \eqref{nestedgeneral} to obtain a full solution of the model.

Due to the free fermionic structure the diagonalization procedure can be done
using standard techniques. The $R$-matrix does not conserve $U(1)$-symmetry, therefore fermion number is not conserved,
and we need to perform a Bogoliubov transformation in this inhomogeneous setting.

Numerical examples are presented in Appendix \ref{app:Y3}.

\subsection{The strategy}

Once the first-level rapidities \(z_1,\dots,z_N\) are fixed, each local nested matrix \(R_{aj}(u,z_j)\) is of
free-fermion type and can be written, up to a scalar prefactor, as a normal-ordered exponential of fermionic
bilinears. After a Jordan-Wigner transformation each local scattering operator is therefore only
quadratic in fermionic creation and annihilation operators.

Because products of fermionic Gaussian operators remain Gaussian, the full nested monodromy matrix is
itself a quadratic-fermion operator. We build the transfer matrix $t(u)$ by taking a partial trace, and once we restrict
$t(u)$ to the fixed parity sectors, we obtain Gaussian operators again.
The spectral problem is therefore reduced from a \(2^N\)-dimensional
many-body diagonalization to a \(2N\)-dimensional one-particle problem.

Let us introduce local fermions acting in the spin spaces $c_j$ and $c_j^\dagger$, $j=1,\dots,N$  using the standard
Jordan-Wigner 
transformation. They satisfy
\begin{equation}
\{c_i,c_j^\dagger\}=\delta_{ij},
\qquad
\{c_i,c_j\}=\{c_i^\dagger,c_j^\dagger\}=0.
\end{equation}
We identify
\begin{equation}
|+\rangle_j \equiv |0\rangle_j,
\qquad
|-\rangle_j \equiv c_j^\dagger |0\rangle_j,
\end{equation}

In order to simplify the explanation  of the strategy, let us dismiss the issue of the fermionic parity for a
moment. Therefore, let us assume that the transfer matrix can be written as
\[
t(u)=\mathcal N(u)\,:\exp\!\Big(\frac12\,\mathcal C^{T}K(u)\mathcal C\Big):,
\]
where we also introduced  the $(2N)$-component vector of operators
\[
\mathcal C=
\begin{pmatrix}
c_1\\ \vdots\\ c_N\\ c_1^\dagger\\ \vdots\\ c_N^\dagger
\end{pmatrix}.
\]

If $W$ is an operator of the form
\[
W=\frac12\,\mathcal C^T K \mathcal C,
\]
then the commutator of \(W\) with any elementary fermion is again linear in the same \(2N\)-dimensional space:
\[
[W,\mathcal C]=A\,\mathcal C
\]
for some \(2N\times 2N\) matrix \(A\). Therefore the Baker--Campbell--Hausdorff formula gives
\[
e^{W}\mathcal C\,e^{-W}=e^{A}\mathcal C.
\]
It follows that the adjoint action of \(t(u)\) closes on the \(2N\)-dimensional one-particle space spanned by
\[
c_1,\dots,c_N,\; c_1^\dagger,\dots,c_N^\dagger.
\]
This is why the full \(2^N\)-dimensional spectral problem reduces to a \(2N\)-dimensional linear problem.

More concretely, let us now define the linear combination of elementary fermions
\[
\Gamma
=\sum_{j=1}^N\bigl(f_j c_j+g_j c_j^\dagger\bigr)
\]
If the coefficients are chosen so that \(\Gamma\) is an eigen-operator of the adjoint action, then
\[
t(u)\,\Gamma\,t(u)^{-1}=\mu(u)\,\Gamma,
\]
or equivalently
\[
t(u)\,\Gamma=\mu(u)\,\Gamma\,t(u).
\]
These \(\Gamma\)'s are the Bogoliubov modes, and the corresponding \(\mu(u)\)'s are the one-particle multipliers.

After diagonalizing the quadratic form, the transfer matrix takes the schematic form
\[
t(u)=\lambda(u)\exp\!\Big(\sum_{i=1}^N \xi_i(u)\,\Gamma_i^\dagger \Gamma_i\Big),
\]
up to a convention-dependent scalar factor absorbed into \(\lambda(u)\). Then the BCH formula immediately implies
\[
t(u)\Gamma_i^\dagger t(u)^{-1}=e^{\xi_i(u)}\Gamma_i^\dagger,
\qquad
t(u)\Gamma_i t(u)^{-1}=e^{-\xi_i(u)}\Gamma_i.
\]
Thus the one-particle multipliers occur in reciprocal pairs,
\[
\mu_i(u)=e^{\xi_i(u)},\qquad \mu_i(u)^{-1}=e^{-\xi_i(u)}.
\]

Let us assume that there is a unique \(\Omega\) vacuum, satisfying
\[
t(u)\Omega=\lambda(u)\Omega,
\qquad
\Gamma_i\Omega=0.
\]
Then for any excited state built by acting with creator modes,
\[
\Gamma_{i_1}^\dagger\cdots \Gamma_{i_n}^\dagger\Omega,
\]
one commutes the transfer matrix through the creators one at a time:
\[
\begin{aligned}
t(u)\,\Gamma_{i_1}^\dagger\cdots \Gamma_{i_n}^\dagger\Omega
&=
\mu_{i_1}(u)\cdots \mu_{i_n}(u)\,
\Gamma_{i_1}^\dagger\cdots \Gamma_{i_n}^\dagger\,t(u)\Omega\\
&=
\lambda(u)\prod_{a=1}^n \mu_{i_a}(u)\,
\Gamma_{i_1}^\dagger\cdots \Gamma_{i_n}^\dagger\Omega.
\end{aligned}
\]
Therefore the transfer-matrix eigenvalues would be
\[
\Lambda_{I}^{\mathrm{tr}}(u)
=
\lambda(u)\prod_{i\in I}\mu_i(u).
\]

This is the transfer-matrix analog of the usual free-fermion Hamiltonian rule. For a quadratic Hamiltonian the energies
are additive, while for the exponential of a quadratic operator the eigenvalues are multiplicative. This is the
conceptual reason why solving the reduced one-particle problem is sufficient to reconstruct the full many-body
spectrum. 

In our case the situation is different, because the transfer matrix is Gaussian only when restricted to the fixed parity
subsectors. Correspondingly, we have two vacua \(\Omega_Q\) with $Q=\pm 1$. However, the  derivations above can be
repeated, with two modifications: every quantity will depend on the parity $Q$ and the only allowed states are those that
have an even number of creation operators on top of the corresponding vacuum.

Let $\Omega_Q$ be the vacuum in parity sector \(Q\), satisfying
\[
t(u)\Omega_Q=\lambda_Q(u)\Omega_Q,
\qquad
\Gamma_i\Omega_Q=0.
\]
Then for any excited state built by acting with an even number of creator modes,
\[
\Gamma_{i_1}^\dagger\cdots \Gamma_{i_n}^\dagger\Omega_Q,
\]
the transfer-matrix eigenvalues become
\[
\Lambda_{Q,I}^{\mathrm{tr}}(u)
=
\lambda_Q(u)\prod_{i\in I}\mu_i(u).
\]
For a set of $N$ elements there are $2^{N-1}$ subsets with an even number of elements, and these even products
can act on two pseudo-vacua. This reproduces the total number of $2^N$ eigenvectors of the transfer matrix.

\subsection{Auxiliary coherent states and the local kernel}

The ordered product in the monodromy matrix acts repeatedly on the same auxiliary space. To convert this ordered product
into a Gaussian integral, we introduce coherent states for the physical and auxiliary fermions. For an overview of this
method, see for example \cite{DupuisFunctionalIntegrals}.

For each physical site \(j\), let
\begin{equation}
|\psi_j\rangle = e^{-\psi_j c_j^\dagger}|0\rangle,
\qquad
\langle \bar\psi_j| = \langle 0| e^{-c_j \bar\psi_j},
\end{equation}
where \(\psi_j,\bar\psi_j\) are Grassmann variables. Likewise, for the auxiliary mode we use
\begin{equation}
|\chi\rangle = e^{-\chi a^\dagger}|0\rangle,
\qquad
\langle \bar\chi| = \langle 0| e^{-a \bar\chi}.
\end{equation}
They satisfy
\begin{equation}
\langle \bar\psi_j | \psi_j \rangle = e^{\bar\psi_j \psi_j},
\qquad
\langle \bar\chi | \chi \rangle = e^{\bar\chi \chi},
\end{equation}
and the usual resolutions of identity,
\begin{equation}
\mathbf 1
=
\int d\bar\psi_j\, d\psi_j\;
e^{-\bar\psi_j \psi_j}\,
|\psi_j\rangle \langle \bar\psi_j|,
\qquad
\mathbf 1_a
=
\int d\bar\chi\, d\chi\;
e^{-\bar\chi \chi}\,
|\chi\rangle \langle \bar\chi|.
\label{eq:fermionic-resolution}
\end{equation}

The reason for introducing \emph{auxiliary} Grassmann variables is the following.  
The monodromy matrix is a product of \(N\) local operators on the same auxiliary line.  
When one inserts \eqref{eq:fermionic-resolution} between two neighboring \(R\)-matrices, one obtains a fresh auxiliary Grassmann pair at that intermediate step. Thus the single auxiliary mode is replaced, in the coherent-state representation, by a chain of Grassmann variables
\begin{equation}
(\bar\chi_0,\chi_0),(\bar\chi_1,\chi_1),\dots,(\bar\chi_{N-1},\chi_{N-1}),
\end{equation}
one pair for each slice of the auxiliary line. These variables are not additional physical degrees of freedom; they are only bookkeeping variables introduced in order to compose the ordered product and perform the auxiliary trace.

The local coherent-state kernel of \(R_{aj}(u,z_j)\) is
\begin{equation}
K_j(\bar\chi_j,\bar\psi_j;\chi_{j-1},\psi_j)
=
\langle \bar\chi_j,\bar\psi_j|R_{aj}(u,z_j)|\chi_{j-1},\psi_j\rangle.
\end{equation}
A direct computation using the $R$-matrix gives
\begin{align}
K_j
&=
a_{+j}
+c_{+j}\,\bar\chi_j\chi_{j-1}
+c_{-j}\,\bar\psi_j\psi_j
+b_j\bigl(\bar\chi_j\psi_j+\bar\psi_j\chi_{j-1}\bigr)
+d_j\bigl(\bar\chi_j\bar\psi_j+\chi_{j-1}\psi_j\bigr)
\nonumber\\
&\hspace{2cm}
+a_{-j}\,\bar\chi_j\bar\psi_j\chi_{j-1}\psi_j,
\label{eq:local-kernel-polynomial}
\end{align}
where
\begin{equation}
a_{\pm j}=a_\pm(u,z_j),\qquad
b_j=b(u,z_j),\qquad
c_{\pm j}=c_\pm(u,z_j),\qquad
d_j=d(u,z_j).
\end{equation}

Introducing
\begin{equation}
x_j=\frac{c_{+j}}{a_{+j}},
\qquad
y_j=\frac{c_{-j}}{a_{+j}},
\qquad
p_j=\frac{b_j}{a_{+j}},
\qquad
q_j=\frac{d_j}{a_{+j}},
\label{eq:local-xy-pq}
\end{equation}
we may write
\begin{equation}
K_j
=
a_{+j}\,
\exp\!\Big(
x_j\,\bar\chi_j\chi_{j-1}
+y_j\,\bar\psi_j\psi_j
+p_j(\bar\chi_j\psi_j+\bar\psi_j\chi_{j-1})
+q_j(\bar\chi_j\bar\psi_j+\chi_{j-1}\psi_j)
\Big).
\label{eq:local-kernel-gaussian}
\end{equation}
This is the local Gaussian building block of the whole construction.
It was the free-fermion identity \eqref{eq:free-fermion-identity} that guaranteed that the quartic term in \eqref{eq:local-kernel-polynomial} is exactly the quadratic completion needed to exponentiate the kernel
into a pure Gaussian.

\subsection{Global Grassmann action}

Fix a parity sector \(Q=\pm1\). In the fermionic transfer-matrix formalism, the restriction to \(\mathcal H_Q\) is implemented by a sector-dependent closure of the auxiliary Grassmann variables,
\begin{equation}
\bar\chi_N=\eta_Q\,\bar\chi_0,
\qquad
\eta_Q=-Q.
\label{eq:closure-sign}
\end{equation}
Thus
\begin{equation}
Q=+1 \ \Longrightarrow\  \eta_{+}=-1
\quad\text{(antiperiodic closure),}
\end{equation}
whereas
\begin{equation}
Q=-1 \ \Longrightarrow\  \eta_{-}=+1
\quad\text{(periodic closure).}
\end{equation}
This is the usual Neveu--Schwarz/Ramond distinction of a free-fermion transfer matrix, now expressed in the language of the auxiliary line.

Let
\begin{equation}
\bar\Psi=(\bar\psi_1,\dots,\bar\psi_N),
\qquad
\Psi=(\psi_1,\dots,\psi_N),
\end{equation}
and assemble all variables into the \(4N\)-component Grassmann vector
\begin{equation}
\Theta=
(\bar\psi_1,\dots,\bar\psi_N,\psi_1,\dots,\psi_N,
 \bar\chi_0,\dots,\bar\chi_{N-1},\chi_0,\dots,\chi_{N-1})^{\mathsf T}.
\label{eq:Theta-ordering}
\end{equation}
Using \eqref{eq:fermionic-resolution}, \eqref{eq:local-kernel-gaussian}, and the closure \eqref{eq:closure-sign}, the transfer-matrix kernel in the sector \(\mathcal H_Q\) takes the form
\begin{equation}
K_Q(\bar\Psi,\Psi)
\propto
\int \prod_{r=0}^{N-1} d\bar\chi_r\, d\chi_r\;
\exp\!\Bigl(\frac12\,\Theta^{\mathsf T}A^{(Q)}\Theta\Bigr),
\label{eq:sector-kernel-full}
\end{equation}
where \(A^{(Q)}\) is a \(4N\times 4N\) antisymmetric matrix.

Equivalently, one may write the exponent itself as
\begin{align}
\frac12\,\Theta^{\mathsf T}A^{(Q)}\Theta
&=
-\sum_{r=0}^{N-1}\bar\chi_r\chi_r
+\sum_{j=1}^{N-1}
\Big[
y_j\,\bar\psi_j\psi_j
+x_j\,\bar\chi_j\chi_{j-1}
+p_j(\bar\chi_j\psi_j+\bar\psi_j\chi_{j-1})
\notag\\
&\hspace{3cm}
+q_j(\bar\chi_j\bar\psi_j+\chi_{j-1}\psi_j)
\Big]
\notag\\
&\quad
+
\Big[
y_N\,\bar\psi_N\psi_N
+\eta_Q x_N\,\bar\chi_0\chi_{N-1}
+\eta_Q p_N\,\bar\chi_0\psi_N
+p_N\,\bar\psi_N\chi_{N-1}
\notag\\
&\hspace{3cm}
+\eta_Q q_N\,\bar\chi_0\bar\psi_N
+q_N\,\chi_{N-1}\psi_N
\Big].
\label{eq:global-action-expanded}
\end{align}
In terms of matrix entries, the nonzero couplings are
\begin{equation}
A^{(Q)}_{\bar\chi_r,\chi_r}=-1,
\qquad r=0,\dots,N-1,
\end{equation}
and for \(j=1,\dots,N\),
\begin{align}
A^{(Q)}_{\bar\psi_j,\psi_j}&=y_j,\\
A^{(Q)}_{\bar\chi_j,\chi_{j-1}}&=x_j,\\
A^{(Q)}_{\bar\chi_j,\psi_j}&=p_j,\\
A^{(Q)}_{\bar\psi_j,\chi_{j-1}}&=p_j,\\
A^{(Q)}_{\bar\chi_j,\bar\psi_j}&=q_j,\\
A^{(Q)}_{\chi_{j-1},\psi_j}&=q_j,
\end{align}
with the convention \(\bar\chi_N=\eta_Q\bar\chi_0\). In particular, every term that contains \(\bar\chi_N\) acquires the factor \(\eta_Q\).

The auxiliary Grassmann variables can now be integrated out exactly.  
Writing \(A^{(Q)}\) in block form according to the decomposition
\begin{equation}
\Theta = (\bar\Psi,\Psi,\bar\chi,\chi),
\end{equation}
namely
\begin{equation}
A^{(Q)}=
\begin{pmatrix}
A^{(Q)}_{\mathrm{phys}} & B^{(Q)}\\
-(B^{(Q)})^{\mathsf T} & D^{(Q)}
\end{pmatrix},
\end{equation}
the Grassmann Schur complement gives
\begin{equation}
\Sigma_Q
=
A^{(Q)}_{\mathrm{phys}}
+
B^{(Q)}\bigl(D^{(Q)}\bigr)^{-1}(B^{(Q)})^{\mathsf T}.
\label{eq:sigma-schur}
\end{equation}
This is a \(2N\times 2N\) antisymmetric matrix, which we split as
\begin{equation}
\Sigma_Q=
\begin{pmatrix}
X_Q & Y_Q\\
-\,Y_Q^{\mathsf T} & Z_Q
\end{pmatrix},
\qquad
X_Q^{\mathsf T}=-X_Q,
\qquad
Z_Q^{\mathsf T}=-Z_Q.
\label{eq:XYZ-blocks}
\end{equation}
After this step the auxiliary variables have disappeared completely. The whole effect of the ordered product and the auxiliary trace is encoded in the \(N\times N\) matrices \(X_Q,Y_Q,Z_Q\).

\subsection{Reduced \(2N\times 2N\) generalized eigenproblem}

The Gaussian operator determined by \(\Sigma_Q\) acts by a Bogoliubov transformation on the physical fermions. We therefore look for operators of the form
\begin{equation}
\Gamma(\bm\xi,\bm\zeta)
=
\sum_{j=1}^N \bigl(\xi_j c_j^\dagger+\zeta_j c_j\bigr),
\qquad
w=
\binom{\bm\xi}{\bm\zeta}\in\mathbb C^{2N},
\label{eq:Bogoliubov-operator}
\end{equation}
satisfying the intertwining relation
\begin{equation}
t(u)\,\Gamma(\bm\xi,\bm\zeta)
=
\nu\,\Gamma(\bm\xi,\bm\zeta)\,t(u)
\qquad\text{on }\mathcal H_Q .
\label{eq:intertwining-relation}
\end{equation}
A standard coherent-state calculation shows that \eqref{eq:intertwining-relation} is equivalent to the \(2N\times 2N\) generalized eigenvalue problem
\begin{equation}
L_Q\,w=\nu\,R_Q\,w,
\label{eq:generalized-problem}
\end{equation}
with
\begin{equation}
L_Q=
\begin{pmatrix}
Y_Q & 0\\
-\,Z_Q & I_N
\end{pmatrix},
\qquad
R_Q=
\begin{pmatrix}
I_N & X_Q\\
0 & Y_Q^{\mathsf T}
\end{pmatrix}.
\label{eq:LqRq}
\end{equation}
This is the reduced problem that replaces the diagonalization of the full \(2^N\times 2^N\) transfer matrix.

The generalized spectrum  consists of \(2N\) roots \(\nu\), which are naturally grouped into \(N\) reciprocal pairs
\begin{equation}
(\nu_1,\nu_1^{-1}),\dots,(\nu_N,\nu_N^{-1}).
\label{eq:reciprocal-pairs}
\end{equation}

\subsection{Selection of the physical one-particle factors}

Every solution \(w\) of \eqref{eq:generalized-problem} defines a Bogoliubov operator \(\Gamma(w)\) through \eqref{eq:Bogoliubov-operator}. Inside each reciprocal pair \((\nu,\nu^{-1})\), one operator acts as a creator on \(\Omega_Q\), whereas the other acts as an annihilator. In practice, one distinguishes them by the vacuum test
\begin{equation}
\Gamma(w)\Omega_Q \neq 0
\quad\Longleftrightarrow\quad
\Gamma(w)\ \text{is a creator},
\end{equation}
while its reciprocal partner satisfies
\begin{equation}
\Gamma(w')\Omega_Q = 0.
\end{equation}
This selects \(N\) physical one-particle factors,
\begin{equation}
\mu_1^{(Q)}(u),\dots,\mu_N^{(Q)}(u),
\label{eq:muq-def}
\end{equation}
one from each reciprocal pair.

The creation/annihilation operators can be identified
by looking at the special point \(u=\infty\) in rapidity space. Indeed, from
\eqref{eq:gaugedRfreefermion} one has
\begin{equation}
\rho(u,z)\to \rho_\infty(z):=-\frac{z^2+1}{z^2-3},
\qquad
\beta(u,z)\to \beta_\infty(z):=-\frac{2z}{z^2-3},
\qquad
\gamma(u,z)=O(u^{-1}),
\end{equation}
while also \(-z/u=O(u^{-1})\). Hence
\begin{equation}
\widetilde R(u,z)=\widetilde R_\infty(z)+O(u^{-1}),
\qquad
\widetilde R_\infty(z)=
\begin{pmatrix}
\rho_\infty(z) & \beta_\infty(z) & 0 & 0\\
0 & 0 & 0 & 0\\
0 & 0 & 0 & 0\\
0 & 0 & \beta_\infty(z) & \rho_\infty(z)
\end{pmatrix}.
\end{equation}
Therefore \(\widetilde t(\infty)\) maps the whole quantum space into
\(\mathrm{span}\{\widetilde A_+,\widetilde A_-\}\), so every non-vacuum eigenvalue
vanishes at \(u=\infty\).

Now let \((\nu,\nu^{-1})\) be one of the reciprocal pairs obtained from the reduced
\(2N\times 2N\) problem. The physical single-particle branch is the one that tends to
zero at infinity:
\begin{equation}
\mu_i^{(Q)}(u)\longrightarrow 0
\qquad (u\to\infty),
\end{equation}
while its reciprocal partner is the annihilator branch. Thus the vacuum is selected
already at \(u=\infty\), without testing the action of the Bogoliubov operators on the
full \(2^N\)-dimensional vector \(\Omega_Q\).

For a numerical computation one only needs the reduced set of equations
\begin{equation}
L_Q w=\nu\,R_Q w,
\end{equation}
or, at generic \(u\) where \(R_Q(u)\) is invertible, the equivalent matrix
\begin{equation}
K_Q(u)=R_Q(u)^{-1}L_Q(u).
\end{equation}
A convenient implementation is to choose two large reference values \(u_1,u_2\) and
diagonalize one generic linear combination
\begin{equation}
K_Q^{\mathrm{ref}}=K_Q(u_1)+\eta\,K_Q(u_2),
\qquad \eta\in\C \ \text{generic},
\end{equation}
which numerically resolves a common mode basis of the commuting family. In that basis
one evaluates \(K_Q(u)\) for the target value of \(u\), pairs the resulting
single-particle factors into \((\nu,\nu^{-1})\), and in each pair keeps the branch that
is small at the large-\(u\) reference point, equivalently the branch that tends to
\(0\) as \(u\to\infty\). Denoting these \(N\) selected branches by
\(\mu_1^{(Q)}(u),\dots,\mu_N^{(Q)}(u)\), the spectrum in the fixed parity sector is
\begin{equation}
\mathrm{Spec}\bigl(t(u)\vert_{\mathcal H_Q}\bigr)
=
\lambda_Q(u)\,
\left\{
\prod_{i\in I}\mu_i^{(Q)}(u)
\;:\;
I\subset\{1,\dots,N\},\ |I|\ \text{even}
\right\}.
\end{equation}
If an accidental degeneracy occurs, one simply changes the reference values
\(u_1,u_2\) (or matches neighboring eigenvectors by overlap along a short path in \(u\)).

It is important to emphasize that \(\Gamma(w)\) changes fermionic parity. Therefore
\begin{equation}
\Gamma(w):\mathcal H_Q \longrightarrow \mathcal H_{-Q}.
\end{equation}
Consequently, the spectrum inside a fixed sector \(\mathcal H_Q\) is generated by \emph{even} products of the selected
one-particle factors.

\bigskip

\vspace{1cm}
{\bf Acknowledgments} 

\bigskip

We are thankful to Tam\'as Gombor, J\'ozsef Konczer, D\'aniel Varga and Timea Vitos Andersson for useful discussions. 
B. P. was supported by the NKFIH excellence grant TKP2021\_NKTA\_64.

\appendix

\section{Numerical tests}

In this Appendix we present numerical solutions of the Bethe Ansatz equations, together with the corresponding energy
levels. Numerical programs to find solutions to the Bethe equations were written by the LLM, in isolated context windows
(without access to the model Hamiltonian). Afterwards, all energy eigenvalues were confirmed by exact
diagonalization. Programs for the latter task were written in part by the human authors, and  in part by the LLM, with
cross-checks in smaller volumes. 

\subsection{Model Y1}

\label{app:Y1}

We chose $\Delta=0.6$ and $L=12$. We focused on selected states with
$N=1, 2, 3, 4$ particles. In each case we present two solutions to the Bethe equations \eqref{Y1BE} and the
corresponding energies as computed from \eqref{Y1E}.

\medskip
$\textbf{N=1}$

\textit{Solution 1.}

$E = 4.8000000000$

$\{p_1\} = \{0\}$

\smallskip
\textit{Solution 2.}

$E = -1.0641016151$

$\{p_1\} = \left\{\frac{\pi}{3}\right\}$

\bigskip
$\textbf{N=2}$

\textit{Solution 1.}

$E = 9.1301425564$

$\{p_1,p_2\} = \left\{-\frac{\pi}{10},\,\frac{\pi}{10}\right\}$

\smallskip
\textit{Solution 2.}

$E = 5.6427384220$

$\{p_1,p_2\} = \left\{-\frac{3\pi}{10},\,\frac{3\pi}{10}\right\}$

\bigskip
$\textbf{N=3}$

\textit{Solution 1.}

$E \approx 16.8529209946$

$\{p_1,p_2,p_3\} \approx \{-1.1249625011,\,-0.5405779536,\,0.0947441279\}$

\smallskip
\textit{Solution 2.}

$E \approx 12.7013598264$

$\{p_1,p_2,p_3\} \approx \{-0.6914066931,\,-0.1314948513,\,0.8229015444\}$

\bigskip
$\textbf{N=4}$

\textit{Solution 1.}

$E \approx 14.6343220591$

$\{p_1,p_2,p_3,p_4\} \approx \{-0.9173734407,\,-0.4025544154,\,1.4186930504,\,1.9956299081\}$

\smallskip
\textit{Solution 2.}

$E \approx 12.7179614457$

$\{p_1,p_2,p_3,p_4\} \approx \{-0.4733695445,\,0.0613164028,\,1.5226678555,\,2.0309779398\}$

\subsection{Model Y2}

\label{app:Y2}

We now list explicit solutions of the nested  Bethe Ansatz equations \eqref{Y2be1}-\eqref{Y2be2} at fixed volume $L=16$
and with a choice $\gamma=0.4$. The
energy is computed from \eqref{Y2E}. 

\bigskip

\textbf{$N=1\quad M=0$}

$E \approx 0.4074098062$

$\{k_1\} \approx \{0.0000000000\}.$

\bigskip
\textbf{$N=2\quad M=0$}

$E \approx -1.4801751651$

$\{k_1,k_2\} \approx \{0.1963495408,\,0.5890486225\}.$

\bigskip
\textbf{$N=2\quad M=1$}

$E \approx -2.9020290100$

$\{k_1,k_2\} \approx \{0.0104665806,\,1.5603297462\}.$

$\{\lambda_1\} \approx \{-0.3592576110 - 0.6100763960\,\mathrm{i}\}.$

\bigskip
\textbf{$N=3\quad M=0$}

$E \approx -2.4889949936$

$\{k_1,k_2,k_3\} \approx \{0.0000000000,\,0.3926990817,\,0.7853981634\}.$

\bigskip
\textbf{$N=3\quad M=1$}

$E \approx -3.0910590845$

$\{k_1,k_2,k_3\} \approx \{-0.1413208872,\,0.2629275700,\,1.4491896441\}.$

$\{\lambda_1\} \approx \{1.4466984681 - 1.2582175984\,\mathrm{i}\}.$

\bigskip
\textbf{$N=4\quad M=0$}

$E \approx -7.7740470628$

$\{k_1,k_2,k_3,k_4\} \approx \{0.1963495408,\,0.5890486225,\,0.9817477042,\,1.3744467859\}.$

\bigskip
\textbf{$N=4\quad M=1$}

$E \approx -6.8452566816$

$\{k_1,k_2,k_3,k_4\} \approx \{0.0168375168,\,0.3901983893,\,1.1667097600,\,1.5678469875\}.$

$\{\lambda_1\} \approx \{-0.3368709948 - 0.6935083965\,\mathrm{i}\}.$

\bigskip
\textbf{$N=4\quad M=2$}

$E \approx -7.7810184447$

$\{k_1,k_2,k_3,k_4\} \approx \{0.2114085104,\,0.5861015162,\,0.9729829331,\,1.3710996939\}.$

$\{\lambda_1,\lambda_2\} \approx \{-0.9287374700 + 1.2403782519\,\mathrm{i},\,-0.4132684172 - 0.4370476056\,\mathrm{i}\}.$

\subsection{Model Y3}

\label{app:Y3}

We now list explicit solutions of the combined Bethe equations at fixed volume $L=16$.  For particle numbers $N=2,
3, 4$ we display a sample solution for selected sectors $Q$ and selected branches of the nested transfer matrix.  

Since this model has an unusual Bethe Ansatz solution, here we summarize the notations and the key formulas, to make a
direct comparison/check straightforward.

For a fixed sector $Q=\pm1$ and a fixed sign pattern $(s_1,\dots,s_N)$ with an even number of minus signs, we solve
the combined Bethe equations
\begin{equation}
z_k^{\,16}\,\Lambda_k=1,
\qquad k=1,\dots,N.
\end{equation}
The branch of the nested transfer matrix is encoded by the convention
\[
s_j=+ \;\Longleftrightarrow\; \text{the fermion }\mu_j\text{ is not added},
\qquad
s_j=- \;\Longleftrightarrow\; \text{the fermion }\mu_j\text{ is added}.
\]
Here $\mu_j$ are one-particle eigenvalues. These eigenvalues come in pairs $\mu_j, \mu_j^{-1}$, and from each pair we
select the one that has a smaller modulus for large rapidity values. This ensures that 
we keep the physical
ones, for which the corresponding fermionic operator does not annihilate the vacuum state. Afterwards, 
the values of the $\mu_j$ are sorted according to their absolute values at a fixed large $u$ value. 

The sector  $Q$ is supplied independently of the sign pattern.  The sign patterns used below
all contain an even number of minus signs.

For each Bethe root $z_k$ we first compute the vacuum transfer-matrix eigenvalue
\[
\lambda_{\mathrm{vac}}(z_k)
=
\prod_{j=1}^N \rho(z_k,z_j)
+
Q\prod_{j=1}^N \beta(z_k,z_j).
\]
Then the transfer-matrix eigenvalue in the chosen branch is
\begin{equation}
\Lambda^{\mathrm{tr}}(z_k)
=
\lambda_{\mathrm{vac}}(z_k)
\prod_{j:\,s_j=-}\mu_j(z_k).
\label{eq:Lambda-tr-branch-example}
\end{equation}
Finally, the quantity entering the Bethe equations is
\begin{equation}
\Lambda_k=-\Lambda^{\mathrm{tr}}(z_k).
\label{eq:Lambda-k-minus}
\end{equation}
Thus for the selected branches used below one has

\[
\begin{array}{ll}
(--,Q=+1):\ \Lambda^{\mathrm{tr}}=\lambda_{\mathrm{vac}}\,\mu_1\mu_2, & (++,Q=-1):\ \Lambda^{\mathrm{tr}}=\lambda_{\mathrm{vac}}, \\[2mm]
(+--,Q=+1):\ \Lambda^{\mathrm{tr}}=\lambda_{\mathrm{vac}}\,\mu_2\mu_3, & (+-+-,Q=+1):\ \Lambda^{\mathrm{tr}}=\lambda_{\mathrm{vac}}\,\mu_2\mu_4, \\[2mm]
(+--+,Q=-1):\ \Lambda^{\mathrm{tr}}=\lambda_{\mathrm{vac}}\,\mu_2\mu_3. &
\end{array}
\]

In the tables we list
the roots $z_k$, the vacuum values $\lambda_{\mathrm{vac}}(z_k)$, the branch values $\Lambda^{\mathrm{tr}}(z_k)$, and separately the
one-particle factors $\mu_j(z_k)$.

The corresponding energies are
computed from 
\[
E=\sum_{j=1}^{N}\bigl(2-z_j^2-z_j^{-2}\bigr).
\]

To make the presentation compact, the rapidities are arranged vertically.  Numerical values are rounded to 6 decimal places.

\paragraph{$N=2$, branch $--$ ($Q=+1$).}
The corresponding energy is
\[
E\approx 1.451675.
\]
The Bethe roots and the one-particle eigenvalues of the transfer matrix are:
\begin{center}
\small
\begin{tabular}{c|c|c|c}
$k$ & $z_k$ & $\lambda_{\mathrm{vac}}(z_k)$ & $\Lambda^{\mathrm{tr}}(z_k)$ \\
\hline
1 & $0.985871+0.167506\,\mathrm{i}$ & $-0.904927+0.425567\,\mathrm{i}$ & $0.900969+0.433884\,\mathrm{i}$ \\
2 & $0.815561+0.578671\,\mathrm{i}$ & $-0.904927-0.425567\,\mathrm{i}$ & $0.900969-0.433884\,\mathrm{i}$ \\
\end{tabular}
\end{center}

\begin{center}
\small
\begin{tabular}{c|c|c}
$k$ & $\mu_1(z_k)$ & $\mu_2(z_k)$ \\
\hline
1 & $-0.677125-0.735868\,\mathrm{i}$ & $0.998113+0.061401\,\mathrm{i}$ \\
2 & $0.677125-0.735868\,\mathrm{i}$ & $-0.998113+0.061401\,\mathrm{i}$ \\
\end{tabular}
\end{center}

Here both signs are minus, hence
\[
\Lambda^{\mathrm{tr}}(z_k)=\lambda_{\mathrm{vac}}(z_k)\mu_1(z_k)\mu_2(z_k),
\qquad
\Lambda_k=-\lambda_{\mathrm{vac}}(z_k)\mu_1(z_k)\mu_2(z_k).
\]

\paragraph{$N=2$, branch $++$ ($Q=-1$).}
The corresponding energy is
\[
E\approx 4.744728.
\]
The Bethe roots and the one-particle eigenvalues of the transfer matrix are:
\begin{center}
\small
\begin{tabular}{c|c|c|c}
$k$ & $z_k$ & $\lambda_{\mathrm{vac}}(z_k)$ & $\Lambda^{\mathrm{tr}}(z_k)$ \\
\hline
1 & $-0.057668-0.998336\,\mathrm{i}$ & $-0.603273-0.797535\,\mathrm{i}$ & $-0.603273-0.797535\,\mathrm{i}$ \\
2 & $0.900274+0.435325\,\mathrm{i}$ & $-0.603273+0.797535\,\mathrm{i}$ & $-0.603273+0.797535\,\mathrm{i}$ \\
\end{tabular}
\end{center}

\begin{center}
\small
\begin{tabular}{c|c|c}
$k$ & $\mu_1(z_k)$ & $\mu_2(z_k)$ \\
\hline
1 & $0.258878+0.965910\,\mathrm{i}$ & $0.988282-0.152639\,\mathrm{i}$ \\
2 & $0.258878-0.965910\,\mathrm{i}$ & $0.988282+0.152639\,\mathrm{i}$ \\
\end{tabular}
\end{center}

Since no excitation is added, one has
\[
\Lambda^{\mathrm{tr}}(z_k)=\lambda_{\mathrm{vac}}(z_k),
\qquad
\Lambda_k=-\lambda_{\mathrm{vac}}(z_k).
\]

\paragraph{$N=3$, branch $+--$ ($Q=+1$).}
The corresponding energy is
\[
E\approx 8.397130.
\]
The Bethe roots and the one-particle eigenvalues of the transfer matrix are:
\begin{center}
\small
\begin{tabular}{c|c|c|c}
$k$ & $z_k$ & $\lambda_{\mathrm{vac}}(z_k)$ & $\Lambda^{\mathrm{tr}}(z_k)$ \\
\hline
1 & $-0.087551-0.996160\,\mathrm{i}$ & $-0.358673+0.933463\,\mathrm{i}$ & $-0.167399-0.985889\,\mathrm{i}$ \\
2 & $0.874428+0.485155\,\mathrm{i}$ & $0.336549+0.941666\,\mathrm{i}$ & $0.248055+0.968746\,\mathrm{i}$ \\
3 & $-0.358369+0.933580\,\mathrm{i}$ & $0.999722-0.023595\,\mathrm{i}$ & $-0.913552-0.406723\,\mathrm{i}$ \\
\end{tabular}
\end{center}

\begin{center}
\small
\begin{tabular}{c|c|c|c}
$k$ & $\mu_1(z_k)$ & $\mu_2(z_k)$ & $\mu_3(z_k)$ \\
\hline
1 & $0.346857+0.937918\,\mathrm{i}$ & $-0.970235+0.242167\,\mathrm{i}$ & $0.958118-0.286373\,\mathrm{i}$ \\
2 & $-0.464927-0.885349\,\mathrm{i}$ & $0.995164-0.098226\,\mathrm{i}$ & $0.981822+0.189803\,\mathrm{i}$ \\
3 & $-0.669121-0.743153\,\mathrm{i}$ & $0.941756+0.336298\,\mathrm{i}$ & $-0.995056-0.099314\,\mathrm{i}$ \\
\end{tabular}
\end{center}

Here the second and third signs are minus, hence
\[
\Lambda^{\mathrm{tr}}(z_k)=\lambda_{\mathrm{vac}}(z_k)\mu_2(z_k)\mu_3(z_k),
\qquad
\Lambda_k=-\lambda_{\mathrm{vac}}(z_k)\mu_2(z_k)\mu_3(z_k).
\]

\paragraph{$N=4$, branch $+-+-$ ($Q=+1$).}
The corresponding energy is
\[
E\approx 4.379447.
\]
The Bethe roots and the one-particle eigenvalues of the transfer matrix are:
\begin{center}
\small
\begin{tabular}{c|c|c|c}
$k$ & $z_k$ & $\lambda_{\mathrm{vac}}(z_k)$ & $\Lambda^{\mathrm{tr}}(z_k)$ \\
\hline
1 & $0.871272-0.490801\,\mathrm{i}$ & $0.167186-0.985925\,\mathrm{i}$ & $0.346800-0.937939\,\mathrm{i}$ \\
2 & $0.789051+0.614328\,\mathrm{i}$ & $0.747713+0.664022\,\mathrm{i}$ & $0.399559-0.916707\,\mathrm{i}$ \\
3 & $0.796048+0.450292\,\mathrm{i}$ & $-11.161978-3.927294\,\mathrm{i}$ & $1.557946+3.871378\,\mathrm{i}$ \\
4 & $-0.951692-0.538334\,\mathrm{i}$ & $-0.079721-0.028049\,\mathrm{i}$ & $0.089461+0.222304\,\mathrm{i}$ \\
\end{tabular}
\end{center}

\begin{center}
\small
\begin{tabular}{c|c|c|c|c}
$k$ & $\mu_1(z_k)$ & $\mu_2(z_k)$ & $\mu_3(z_k)$ & $\mu_4(z_k)$ \\
\hline
1 & $0.457667+0.889124\,\mathrm{i}$ & $0.969564+0.244837\,\mathrm{i}$ & $0.978083+0.208216\,\mathrm{i}$ & $0.998130-0.061131\,\mathrm{i}$ \\
2 & $0.900098-0.435687\,\mathrm{i}$ & $-0.347904-0.937530\,\mathrm{i}$ & $-0.989861-0.142043\,\mathrm{i}$ & $0.999193+0.040175\,\mathrm{i}$ \\
3 & $1.023390-0.341770\,\mathrm{i}$ & $-0.238114-0.265322\,\mathrm{i}$ & $0.055624+0.312523\,\mathrm{i}$ & $0.989208+0.010379\,\mathrm{i}$ \\
4 & $-0.879100+0.293583\,\mathrm{i}$ & $1.873527+2.087603\,\mathrm{i}$ & $-0.552014-3.101512\,\mathrm{i}$ & $-1.010799-0.010605\,\mathrm{i}$ \\
\end{tabular}
\end{center}

Here the second and fourth signs are minus, hence
\[
\Lambda^{\mathrm{tr}}(z_k)=\lambda_{\mathrm{vac}}(z_k)\mu_2(z_k)\mu_4(z_k),
\qquad
\Lambda_k=-\lambda_{\mathrm{vac}}(z_k)\mu_2(z_k)\mu_4(z_k).
\]

\paragraph{$N=4$, branch $+--+$ ($Q=-1$).}
The corresponding energy is
\[
E\approx 8.690837.
\]
The Bethe roots and the one-particle eigenvalues of the transfer matrix are:
\begin{center}
\small
\begin{tabular}{c|c|c|c}
$k$ & $z_k$ & $\lambda_{\mathrm{vac}}(z_k)$ & $\Lambda^{\mathrm{tr}}(z_k)$ \\
\hline
1 & $-0.896620-0.442801\,\mathrm{i}$ & $0.586203+0.810164\,\mathrm{i}$ & $-0.492064+0.870559\,\mathrm{i}$ \\
2 & $0.872201-0.489147\,\mathrm{i}$ & $0.968078-0.250648\,\mathrm{i}$ & $0.318174-0.948032\,\mathrm{i}$ \\
3 & $-0.084685-0.996408\,\mathrm{i}$ & $0.407952+0.913003\,\mathrm{i}$ & $-0.212575-0.977145\,\mathrm{i}$ \\
4 & $0.505427-0.862869\,\mathrm{i}$ & $-0.267572-0.963538\,\mathrm{i}$ & $0.584328+0.811517\,\mathrm{i}$ \\
\end{tabular}
\end{center}

\begin{center}
\small
\begin{tabular}{c|c|c|c|c}
$k$ & $\mu_1(z_k)$ & $\mu_2(z_k)$ & $\mu_3(z_k)$ & $\mu_4(z_k)$ \\
\hline
1 & $-0.929255+0.369440\,\mathrm{i}$ & $-0.261967-0.965077\,\mathrm{i}$ & $-0.986433+0.164167\,\mathrm{i}$ & $-0.999972+0.007438\,\mathrm{i}$ \\
2 & $0.809926+0.586532\,\mathrm{i}$ & $0.976160+0.217051\,\mathrm{i}$ & $0.350739-0.936473\,\mathrm{i}$ & $0.996455-0.084125\,\mathrm{i}$ \\
3 & $-0.178942+0.983860\,\mathrm{i}$ & $0.520584+0.853810\,\mathrm{i}$ & $-0.684221+0.729275\,\mathrm{i}$ & $-0.720590-0.693362\,\mathrm{i}$ \\
4 & $0.415302+0.909683\,\mathrm{i}$ & $0.828819+0.559517\,\mathrm{i}$ & $-0.584135+0.811656\,\mathrm{i}$ & $0.781031-0.624493\,\mathrm{i}$ \\
\end{tabular}
\end{center}

Here the second and third signs are minus, hence
\[
\Lambda^{\mathrm{tr}}(z_k)=\lambda_{\mathrm{vac}}(z_k)\mu_2(z_k)\mu_3(z_k),
\qquad
\Lambda_k=-\lambda_{\mathrm{vac}}(z_k)\mu_2(z_k)\mu_3(z_k).
\]


\providecommand{\href}[2]{#2}\begingroup\raggedright\endgroup

\end{document}